\documentclass[nofootinbib,superscriptaddress]{revtex4}
\usepackage{graphicx,epsfig}
\usepackage{amsmath}
\usepackage{amsfonts}
\usepackage{braket}
\usepackage{fancyhdr}
\usepackage{color}
\usepackage{comment}
\newcommand{\be}{\begin{eqnarray}}
\newcommand{\ee}{\end{eqnarray}}

\newcommand{\xo}{\vec{x}_0}
\newcommand{\g}{g_R(\vec{x}_0)}
\newcommand{\C}{{\cal C}}

\includecomment{details}
\excludecomment{usual}

\begin{document}
	
	\title{Non-linear statistics of primordial black holes from gaussian curvature perturbations}
	\author{Cristiano Germani}
	\email{germani@icc.ub.edu}
	\affiliation{Institut de Ci\`encies del Cosmos, Universitat de Barcelona, Mart\'i i Franqu\`es 1, 08028 Barcelona, Spain}
	\author{Ravi K. Sheth}
	\email{shethrk@upenn.edu}
	\affiliation{Center for Particle Cosmology, University of Pennsylvania, Philadelphia, PA 19104, USA}
		
	\begin{abstract}
		We develop the non-linear statistics of primordial black holes generated by a gaussian spectrum of primordial curvature perturbations. This is done by employing the compaction function as the main statistical variable under the constraints that: a) the over-density has a high peak at a point $\vec{x}_0$, b) the compaction function has a maximum at a smoothing scale $R$, and finally, c) the compaction function amplitude at its maximum is higher than the threshold necessary to trigger a gravitational collapse into a black hole of the initial over-density. Our calculation allows for the fact that the patches which are destined to form PBHs may have a variety of profile shapes and sizes.  The predicted PBH abundances depend on the power spectrum of primordial fluctuations. For a very peaked power spectrum, our non-linear statistics, the one based on the linear over-density and the one based on the use of curvature perturbations, all predict a narrow distribution of PBH masses and comparable abundance. For broader power spectra the linear over-density statistics over-estimate the abundance of primordial black holes while the curvature-based approach under-estimates it. Additionally, for very large smoothing scales, the abundance is no longer dominated by the contribution of a mean over-density but rather by the whole statistical realisations of it.
	\end{abstract}
	
	\maketitle

\section{Introduction}
Primordial Black Holes (PBHs) are the most economical option for explaining Dark Matter (DM). Indeed, if generated by large fluctuations of scalar primordial perturbations, full explanation of DM in terms of PBHs only depends on a thorough understanding of inflation. Observationally, PBHs could account for some and/or all of the DM if they are in the mass range $\left[10^{-16},10^{-10}\right]\, M_\odot$ \cite{last}.  Detection of such low mass black holes would be a definitive proof of their primordial, rather than astrophysical, origin and might also be a non-trivial test for the inflationary paradigm.

In scenarios in which the formation of a PBH is triggered by a large initial curvature perturbation generated by the inflaton, the gravitational collapse into a black hole is an extremely rare event which occurs during the radiation-dominated epoch. Nevertheless, once formed, such black holes behave like dust particles so their density grows (by one power of the scale factor) with respect to that of the background radiation. Thus, only a tiny fraction of the total energy density at the time of PBH formation is needed to match the required abundance of DM today.

Although PBHs are, by definition, large non-linear over-densities, if formed from rare perturbations in the initial field, their statistics might still be fully inferred by the use of perturbation theory.  

Inflation provides the statistical distribution of curvature perturbations. Single field scenarios are typically characterised by an approximately gaussian statistics of curvature perturbations.  Therefore, as a simplifying assumption, we shall assume in this paper exact gaussianity (discussions about this assumption in specific models of inflation can be found e.g. in \cite{vicente, vennin}).

 The linear relation between curvature perturbations and over-densities has been used in \cite{meilia} to calculate the abundance of PBHs via the statistics of high (gaussian) peaks \cite{bbks} of over-densities. This was done under the assumption that the non-linearities, relating the over-density to curvature perturbations, should be statistically irrelevant in the distributions of the peak amplitudes, given the fact that they are proportional to higher correlators of the curvature perturbations. This approximation might not be correct in the large fluctuation tails which lead to PBH formation. Moreover, this approach has a second drawback. The threshold for PBH formation, usually given in terms of the gravitational energy, was mapped into a threshold in the over-density amplitude. Both the threshold and the mapping depend upon the spatial profile of the fluctuation \cite{musco} thus, different statistical realizations of different high peaks would correspond to different thresholds. Nevertheless, earlier works have assumed that the dispersion of these would not be large and considered only the mean threshold corresponding to the mean high peak profile. 

To try to overcome the first problem, the Authors in \cite{jaume} worked directly with curvature perturbations assuming their high-peaks statistics to coincide with the ones of over-densities. However, over-density maxima might also be related to minima of local curvature, especially above threshold where the PBH masses are not vanishing.  This could result in underestimating the true abundance. 

The linear theory approach to PBHs abundance used in \cite{meilia} has been questioned in \cite{riotto}. There, the authors have considered the role of the skewness obtained from the non-linear relation between the over-density and the (gaussian) curvature perturbations. Those non-gaussianities are typically very small with respect to the two-point functions, thus one would be tempted to discard them \cite{meilia}. However, as already mentioned, the statistics of PBHs is related to the (exponentially suppressed) tail area of the over-density probability distribution. Thus, small deviations to gaussianity might nevertheless give important contributions \cite{ineludible,vennin}. Indeed, \cite{riotto} have found that the skewness contribution to the mean density profile is of the same order as the one obtained from linear theory. This would immediately imply the failure of the perturbative approach in terms of momenta, and so, the failure of the linear approximation. Interestingly though, the skewness contribution does not substantially change the shape of the over-density and thus acts just as a ``renormalization'' of the over-density amplitude. This has led \cite{riotto} to suspect that the re-summation of the curvature momenta in the over-density might not change substantially the linearly calculated abundance of PBHs. A similar conclusion has been drawn in \cite{ineludible,chris} where it has been argued that non-linear corrections to the relation between over-densities and curvature perturbation may only slightly change the required amplitude of the curvature power spectrum. We will confirm this expectation for a very narrow power spectrum while confuting it in the broad case.

Finally, a further approach aiming to take into account the non-linearities in the calculation of the PBHs abundance has been considered in \cite{chris}. At the full non-linear level, the PBH mass is a function of the averaged {\it linear} over-density $\delta_\ell$ on a sphere of radius $R$. Here, $R$ is the maximum radius of the gravitational potential (or more technically the ``compaction function'' \cite{compaction}) associated to the {\it non-linear} over-density $\delta$. More specifically, $\delta_\ell$ is the averaged over-density calculated via its linear relation to curvature perturbations. In \cite{chris}, the authors used the probability distribution of rare peaks in $\delta_\ell$ (i.e. using the methods of \cite{meilia}) to estimate PBH abundances. Although this approach ameliorates the linear approach of \cite{meilia}, it still suffers from the same main problems. Knowing the probability distribution of rare peaks in $\delta_\ell$ is not enough: in order to perform the averaging of the over-density, the (statistical) condition that $R$ is the compaction function maximum must be imposed. Moreover, the predicted PBH abundance depends exponentially on the critical value for $\delta_\ell$. This critical value depends upon the full compaction function profile \cite{musco} or at least, to a very good approximation, upon a specific combination of the amplitude and second derivative of the compaction function at its maximum \cite{universal}. Thus, knowledge of $\delta_\ell$ peaks statistics alone is not enough to estimate PBHs abundances.  

In what follows, we improve on previous works in three ways:
\begin{itemize}
  \item First, we impose the additional condition that $R$ is a maximum of the compaction function.
  \item Second, thanks to the remarkable results of \cite{universal}, we include the fact that the threshold value depends on the compaction function shape directly in the statistics.
  \item Third, in our estimate of the PBH abundance, $R$ is a free parameter rather than a single fixed scale related to an assumed {\it mean} compaction function. In essence, this is because we go beyond the `peaks theory' of \cite{bbks} and instead use methods based on the `excursion set peaks theory' of \cite{esp}.
\end{itemize}
As a result, our approach considers all possible $R$ and, given $R$, we account for all allowed compaction function shapes.
Thus, for the first time, we are able to consider the whole mass spectrum and thus, the cumulative abundance of the PBHs generated from the radiation epoch to equality via large and rare inflationary perturbations of over-densities. 

\section{Variable definitions}

At super-horizon scales and at leading order in gradient expansion, a perturbation on a Friedman-Robertson-Walker (FRW) spacetime is well described by the following metric: 
\be
 ds^2 = -dt^2 + a(t)^2 e^{2\zeta({\vec x})}d\vec{x}\cdot d\vec{x}\ ,
\ee   
and associated radiation over-density $\delta\rho(\vec{x},t)$
\be\label{full}
\frac{\delta\rho(\vec{x},t)}{\rho(t)}=-\frac{8}{9}\frac{1}{a(t)^2H(t)^2}e^{-\frac{5}{2}\zeta(\vec{x})}\nabla^2 \left(e^{\frac{\zeta(\vec{x})}{2}}\right)\ ,
\ee
where $\nabla^2$ is the flat space Laplacian, $H=\dot a/a$ the Hubble constant and $\rho(t)$ is the background density. 

In what follows, it is useful to work with the Fourier modes of the field $\zeta$, which we call $\zeta({\vec{k}})$.  These $k$-modes define a field in position/coordinate space such that, at any position $\vec{x}$, the value of the field is given by 
\be
 \zeta(\vec{x}) \equiv \int \frac{d\vec{k}}{(2\pi)^3}\,e^{i{\vec{k}}\cdot{\vec{x}}}
 \zeta(\vec{k}) 
 \qquad{\rm and\ so}\qquad
 \nabla^2\zeta(\vec{x}) = -\int \frac{d\vec{k}}{(2\pi)^3}\,e^{i{\vec{k}}\cdot{\vec{x}}}\,k^2\zeta(\vec{k}).
\ee
Expanding the exponentials in Eq.(\ref{full}) we get 
\be
\frac{\delta\rho(\vec{x},t)}{\rho(t)} \approx
-\frac{8}{9}\frac{1}{a(t)^2H(t)^2}\,\left[1-\frac{5}{2}\zeta(\vec{x})\right]\frac{\nabla^2\zeta(\vec{x})}{2}
\approx -\frac{4}{9}\frac{\nabla^2\zeta(\vec{x})}{a(t)^2H(t)^2}
= \frac{4}{9}\frac{1}{a(t)^2H(t)^2}\int \frac{d\vec{k}}{(2\pi)^3}\,e^{i{\vec{k}}\cdot{\vec{x}}}\,k^2\zeta(\vec{k}).
\ee
We refer to the quantity on the right hand side as the linear overdensity, since we linearized Eq.(\ref{full}).  The Fourier transform of this linear overdensity is 
\be
 \Delta(\vec{k}) =  \frac{4}{9}\frac{k^2\zeta(\vec{k})}{a(t)^2H(t)^2}.
\ee
The linear overdensity averaged within a sphere of radius $R$ centered at $\xo$ satisfies 
\begin{align}
 \delta_R(\xo,t)
  &\equiv \frac{3}{4\pi R^3}\int d\vec{x}\,
        \frac{\delta\rho(\vec{x},t)}{\rho(t)}\,
        \theta\left(R-\Big|\vec{x}-\xo\Big|\right)\\   
  &= \int \frac{d\vec{k}}{(2\pi)^3}\,\Delta(\vec{k},t)\,
            e^{i{\vec{k}}\cdot{\xo}}\,W_{\rm TH}(kR) ,
\end{align}
where
\be 
 W_{\rm TH}(kR)\equiv 3\,\frac{j_1(kR)}{kR}
                  = 3\,\frac{\sin(kR) - kR\, \cos(kR)}{(kR)^3}\  ,
\ee
is the Fourier transformed top-hat Window function.

The non-linear over-density represents the physical observable which defines the local spacetime deformation from a pure FRW universe. A PBH is the direct consequence of the gravitational collapse of such a perturbation. However, the explicit use of the over-density is actually un-necessary. The main statistical variable, directly related to inflation, is in fact $\zeta$, as pointed out in \cite{jaume}.  Nevertheless, we will still sometimes use the linearized overdensity -- in terms of $\zeta$ -- for comparison to previous work \cite{meilia,riotto}. 

The quantity which plays a crucial role in the physics of PBH formation is closely related to the gravitational potential and, in spherical symmetry, is known as the compaction function \cite{compaction}.  For a spherically symmetric over-density distribution the compaction function ${\cal C}$ is a particular nonlinear combination of the smoothed overdensity \cite{yoo}:
\be\label{CS}
 {\cal C}(R,\xo)=\frac{R^2}{R_H^2}\delta_R(\xo,t)
 \left(1-\frac{3}{8} \frac{R^2}{R_H^2}\delta_R(\xo,t)\right)\ .
\ee
Note that $\C$ is independent of time $t$ because the time dependence of $\delta_R$ is canceled by that of $R_H^2$.  Given the central importance of the combination $(R^2/R_H^2)\,\delta_R(\xo,t)$ in the compaction function, we define the new gaussian variable 
\be
 g_R(\xo) \equiv \frac{R^2}{R_H^2}\delta_R(\xo,t)\
   = \frac{4}{9}\int \frac{d\vec{k}}{(2\pi)^3}\,
     (kR)^2\zeta(\vec{k})\,W_{\rm TH}(kR) \,e^{i{\vec{k}}\cdot{\xo}},
\ee
which is explicitly independent of time.  In terms of $g_R$,
\be\label{lastmin}
 {\cal C}(R,\xo) = g_R(\xo)\left(1 - \frac{3}{8} g_R(\xo)\right)\ ;
 \label{Candg}
\ee
this shows that $\C\leq 2/3$, with equality at $\g=4/3$. Our goal is to determine what is required of $g_R(\xo)$ so that a black hole forms around it.

\section{Black hole formation: conditions and masses}

\subsection{Conditions}

As we have already mentioned, numerical simulations have shown that for a black hole to form at position $\xo$, the compaction function should be maximal on some scale $R_{\rm max}$ (from now on simply $R$) and its amplitude on this scale should exceed a critical value ${\cal C}_c$ \cite{compaction}.  This threshold value depends upon the whole curvature profile \cite{musco}. In particular, simulations show the existence of a lower bound ${\cal C}_c\geq 2/5$ \cite{universal}. This latter, combined with the upper bound $\C\leq 2/3$, implies that 
\be\label{limits}
\frac{4}{3}\left(1 - \sqrt{2\over 5}\right)\simeq 0.49\leq g_R^c(\xo)\equiv g_c\leq \frac{4}{3}\quad\cup\quad
\frac{4}{3}\leq g_c\leq \frac{4}{3}\left(1 + \sqrt{2\over 5}\right)\simeq 2.18\ . 
\ee
To within a few percent, ${\cal C}_c$ only depends on $R^2{\cal C}''(R,\xo)\Big|_{\C=\C_c}\equiv R^2\C''_c(R,\xo)$ via the analytical transcendental equation \cite{universal}
\be\label{deltac}
 {\cal C}_c = \frac{4}{15} e^{-\frac{1}{q}}\frac{q^{1-\frac{5}{2q}}}{\Gamma\left(\frac{5}{2q}\right)-\Gamma\left(\frac{5}{2q},\frac{1}{q}\right)}\ ,
\ee
where
\be\label{q}
 q\equiv-\frac{{\cal C}_c''(R,\xo)R^2}{4\,{\cal C}_c(R,\xo)}\ ,
\ee
and where we have used the notation $df/dR\equiv f'$.

We need now to find the conditions for $R$ to be the maximum of $\C$. The first condition is that the compaction function has an extreme in $R$, i.e. $\C' = g_R' (1 - \frac{3}{4}g_R) = 0$. The latter implies either $g_R'=0$ or $g_R= 4/3$. The value $g_R=4/3$ corresponds to the upper bound of $\C$ and so it is statistically unluckily. Discarding it, we are left with the condition $g'_R(\vec{x}_0)=0$. The second requirement for $R$ to be a maximum of the compaction function is that $\C'' = g_R''(1 - \frac{3}{4} g_R)<0$. This directly implies that a maximum of $\C$ corresponds to a maximum of $\g$ for $0.49\leq \g < 4/3$ and to a minimum for $4/3 < \g\leq 2.18$. 

Because we are looking for isolated peaks in the over-density, we will also need that $\xo$, the position around which $\C'=0$, is a local peak in $\C(R,\xo)$. This is achieved by demanding that $\vec{\nabla}_{\xo}\C(R,xo)=0$, and that the $3\times 3$ matrix of second derivatives of $\C$ be negative. As before, a maximum of $\C$ in $\xo$ corresponds to a maximum of $\g$ for $0.49\leq \g < 4/3$ and to a minimum for $4/3 < \g\leq 2.18$. 

Strictly speaking, if there is a peak at position $\xo$ when the smoothing scale is $R$, then a small change in $R$ may result in a slightly shifted peak.  To impose the spatial peak constraint across smoothing scales (i.e. along the path traced by the shifting peak position as $R$ changes), we must impose the peak requirement slightly differently.  We discuss this in Appendix~\ref{x2p}, where we provide the exact expression, and argue that ignoring this extra subtlety should be a good approximation for the high peaks of most interest here.

Summarizing then, black hole formation is associated with those positions $\xo$ for which
\begin{itemize}
	\item Maximum of $\g$ in $R$: $0.49\leq g_R < 4/3$
	\be
  g_R'=0, \quad  g_R''< 0,
  \quad \nabla_i g_R=0, \quad {\rm and}\quad \nabla_i\nabla_j g_R<0\ .
 \label{conditions}
\ee
\item Minimum of $\g$ in $R$: $4/3< g_R \leq 2.18$
	\be
g_R'=0, \quad  g_R''> 0,
\quad \nabla_i g_R=0, \quad {\rm and}\quad \nabla_i\nabla_j g_R>0\ ,
\label{conditions2}
\ee
\end{itemize}
where $\nabla_i$ is the derivative with respect to $x_0^i$.

We pause here to stress that while the condition to form a PBH is related to the presence of a maximum in the compaction function, PBHs are associated with maxima of the overdensity (Eq.~\ref{conditions}) or minima (Eq.~\ref{conditions2}), depending on the value of $\g$.  While the dependence of the PBH abundance on over-density maxima have been considered in the literature \cite{meilia,chris,riotto}, the minima have been overlooked. In the next section, we estimate the number density of such positions and show that the minima may indeed contribute non-negligibly.

In what follows, we work with the dimensionless variables
\be
 v_R&\equiv& R\,g_R'\ ,\cr
 w_R &\equiv& -R^2\,g_R''\ ,\cr 
 \chi_R &\equiv& -R^2\,\nabla^2g_R\ . 
\ee
Then $\g$ can then be related to the others by the relation 
\be \label{relation}
 R^2g_R'' - R^2\nabla^2 g_R = 2g_R \ .
\ee
Therefore, given any two of the variables $g,w$ and $\chi$, the third is determined by the relation
\be
 \chi_R = 2g_R + w_R\ .
\ee
The last equation might also be read as showing the relation between different smoothings of $\Delta$: $g_R$ is the smoothing of $(R/R_H)^2\,\Delta(k)$ via the top-hat window function $W_{\rm TH}(kR)$, whereas $\chi_R$ and $w_R$ smooth it with $(kR)^2\,W_{\rm TH}(kR)$ and $[(kR)^2-1]\,W_{\rm TH}(kR)$ respectively.

To simplify the notation from now on we will drop the subscript $R$ in our statistical variables, unless needed.

 \subsection{Masses}
 
So far, we have discussed what is special about the positions around which PBHs form.  For $\C$ greater than but close to the critical value $\C_c$ (c.f. equation~\ref{deltac}), the mass of the associated PBH is described by the following scaling law \cite{scaling}: \footnote{Larger peaks deviate from this scaling law by $15\%$ \cite{albert}, however they are also exponentially rarer. Therefore, we neglect the fact that this scaling law may be violated.}
\be
 M_\bullet = \frac{{\cal K}}{2 H_R}\left(\C - \C_c\right)^\gamma,
  \qquad{\rm with}\qquad \gamma\approx 0.36,
 \label{Mbh}
  \ee
where $\C_c=\C_c(w)$ (see eq. \eqref{deltac}) and $\C$ is given by eq. \eqref{lastmin}.
 
We have also identified $H_R^{-1}(t_m) = a(t_m)R$ \cite{musco} which can be inverted to obtain $t_m(R)$ during radiation domination, given the fact that $a(t)=a_0\left(1+2 H_0 (t-t_0)\right)^{1/2}$, where $a_0$ and $H_0$ are respectively the scale factor and the Hubble scale related to some reference time $t_0$. 

Note that $t_m(R)$ is only a reference time relating the smoothing scale $R$ with the mass contained in a Hubble volume of the same size. In other words, although recently called ``horizon crossing'' time in \cite{musco}, physically, $t_m(R)$ is generically unrelated with the time in which the over-density, eventually collapsing into a PBH, re-enter the Hubble scale. The latter is determined by the full non-linear evolution of the collapsing perturbation. 
 
Finally, the exact value of the constant ${\cal K}$ depends on the full dynamics of the PBH accretion.  However, it is always of ${\cal O}(1)$ \cite{musco}. For a gaussian profile in the local curvature, one has ${\cal K}\simeq 6$ \cite{albert}. As this constant will not change the order of magnitude of our estimations for the abundance, we will fix it to ${\cal K}\sim 6$ whenever needed.

Gravitational collapse into a non vanishing mass PBH happens whenever $\C(w,\chi)>\C_c(w)$, implying
\be
\frac{4}{3}\left(1-\sqrt{1-\frac{3}{2} \C_c(w)}\right)< g< \frac{4}{3}\left(1+\sqrt{1-\frac{3}{2} \C_c(w)}\right)\ .
\ee
Although gravitational collapse into a PBH can happen for any $g>\frac{4}{3}\left(1-\sqrt{1-\frac{3}{2} \C_c(w)}\right)$ in linear theory \cite{meilia}, non-linearities put an upper bound on the allowed range of linear amplitudes. 

\section{Statistics}

As we have already mentioned, the statistical proprieties of all the variables introduced so far are fully determined by the statistics of $\zeta$. Using ${\cal P}(k)$ to denote the power spectrum of $\zeta_k$ one has 
\be
\langle\zeta_k\zeta_{k'}\rangle\equiv (2\pi)^3\times\frac{2\pi^2{\cal P}(k)}{k^3}\delta^{(3)}\left(k+k'\right)\ .
\ee
If $\zeta_k$ is a gaussian variable, then $\Delta$ (and hence $\delta_R$) is also.  Its Fourier transform $\Delta_k$ has variance 
\be\label{sig}
 \langle\Delta_k\Delta_{k'}\rangle \equiv (2\pi)^3\times R_H^4\times\frac{16}{81}\,k\times 2\pi^2{\cal P}(k)\,\delta^{(3)}(k+k')\ .
\ee
Defining 
\be
 \sigma_j^2(R) \equiv \frac{16}{81}\int \frac{dk}{k}\,(kR)^4\,{\cal P}(k)\,W_{TH}^2(kR)\,(kR)^{2j}\ ,
\ee
and using \eqref{relation}, we have the following relevant correlators:
\be
 \sigma_g^2 &\equiv& \langle g^2\rangle = \sigma_0^2,\qquad
 \sigma_\chi^2 \equiv \langle \chi^2\rangle = \sigma_2^2\ ,\qquad
 \sigma_w^2 \equiv \langle w^2\rangle = \sigma_2^2 - 4\sigma_1^2 + 4\sigma_0^2,
 \cr
 \sigma_v^2 &\equiv& \langle v^2\rangle
 = \frac{d\langle gv\rangle}{d\ln R} - \langle gv\rangle + \sigma_1^2 - 2\sigma_0^2, \qquad {\rm where}\qquad
 \langle gv\rangle=\frac{1}{2}\frac{d\sigma_0^2}{d\ln R}
\ee
 and
\be
 \label{gammas}
 \langle g\chi\rangle&=&\sigma_1^2\ ,\qquad
 \langle gw\rangle=\sigma_1^2-2\sigma_0^2\ ,\qquad
 \langle w\chi\rangle=\sigma_2^2-2\sigma_1^2\ ,\qquad
 \cr
 \langle v \chi\rangle&=&\frac{1}{2}\frac{d\sigma_1^2}{d\ln R}-\sigma_1^2\ ,
 \qquad
 \langle v w\rangle=\langle v \chi\rangle-2\langle v g\rangle.
\ee
Finally, it is standard to define the normalized (Pearson) correlation coefficient $\gamma_{ab}\equiv \langle ab\rangle/\sigma_a\sigma_b$.  Unless ${\cal P}$ is a power law, these $\gamma_{ab}$ also depend on $R$.
 
\subsection{Implementing the constraints: number density}

We will start implementing the constraints related to a maximum of $g$ (Eq.~\ref{conditions}). A similar procedure can be straightforwardly applied to the minima.

The constraints on $g$ are
\be
 \frac{4}{3}\left(1-\sqrt{1-\frac{3}{2} \C_c(w)}\right)< g< \frac{4}{3}\ ,
\ee
we stress once again that this implies $\C>\C_c(w)$.
The logic for implementing the constraints on the derivatives of $g$ with respect to the smoothing scale $R$ is analogous to that leading to Eq.3.5 in Section~III$b$ of \cite{bbks}. We require that $dg_R/dR > 0$ and $dg_{R+\Delta R}/dR<0$ as $\Delta R\to 0$. Since $dg_{R+\Delta R}/dR \approx g_R' + \Delta R\,g_R''$, this means we want $0 < g_R' < -\Delta R\,g_R''$ and $g_R''\le 0$. In our dimensionless variables we then impose
\be
 0 < v < (\Delta R/R)\,w\ ,\ \mbox{and}\ ,\ w\ge 0\ .
\ee 
A similar argument applies to how one should implement the spatial peak constraints on $\nabla_i\nabla_j \g$ and $\chi$ (see Section IIIa of \cite{bbks} for details).  
 
Therefore, in the approximation in which peaks in $\C$ are peaks in $g$, the comoving number density of positions which satisfy the conditions~(\ref{conditions}) and which produce PBHs of mass $M_\bullet$ is given by
\be
 \!\!\!\!\!\frac{dn}{dM_\bullet} =\int_{R_{\rm min}}^{R_{\rm max}} \frac{dR}{R}\int_0^\infty dw\, w\, \int_{\frac{4}{3}\left(1-\sqrt{1-\frac{3}{2} \C_c(w)}\right)}^{4/3} dg\,  \frac{f(\chi/\sigma_\chi)}{(2\pi R_*^2)^{3/2}}\,p(g,w,v=0)\,
 \delta_{\rm D}\left(M_\bullet - \frac{{\cal K}}{2 H_R}\,\Bigl[\C(g) - \C_c(w)\Bigr]^{0.36}\right),\nonumber
\ee
(see appendix \ref{x2p}, where we discuss the approximation which leads to this expression), where $R_{\rm min}\approx 10^{12}$~gr (corresponding to small black holes that would have been Hawking evaporated by equality),
$R_{\rm max}\approx R_{eq}$ (the Hubble radius at matter-radiation equality) and
$R_*\equiv R\sqrt{3}\sigma_1/\sigma_2$.
The terms involving $f(\chi/\sigma_\chi)/(2\pi R_*^2)^{3/2}$ come from the spatial peaks constraint developed in~\cite{bbks} with $f(x)$ given by their Eq.~(A15); if $x\gg 1$ one finds $f(x)\to x^3$. The function $p(g,w,v=0)$ denotes the joint distribution of the variables $g,v$ and $w$, for which we provide an explicit expression shortly.  

The delta function selects only those combinations of $R$, $g$ and $w$ which return $M_\bullet$ (c.f. the scaling relation Eq.~\ref{Mbh}). Note that, the integrals over $R$, $g$ and $w$ represent contributions to the PBH abundance from different possible profile shapes and sizes (the shape explicitly determines the critical value of the compaction function $\C_c$).  
In this constraint, $\C$ being a function of only $g$ follows directly from Eq.~(\ref{Candg}).  However, the fact that $\C_c$ is a function of only $w$ deserves further comment.  First, recall that $-R^2\C'' = -R^2g'' (1 - 3g/4) = w\, (1 - 3g/4)$.  Next, note that, for $g<4/3$, $(1-3g/4) = \sqrt{1 - 3\C/2}$, so $q$ of Eq.~(\ref{q}) equals $-R^2\C_c''/4\C_c = w\, \sqrt{1 - 3\C_c/2}/4\C_c$.  Hence, Eq.~(\ref{deltac}) can be solved (numerically) to yield $\C_c$ as a function of $w$.  The result is well-approximated by
\be
 {\cal C}_c(w) \approx  
    \frac{2}{5} + \frac{4}{15}\,\frac{1 + {\rm erf}(\omega/\sqrt{2})}{2},\qquad
    {\rm where}\quad \omega \equiv \frac{\ln(|w|/10.41)}{2.45}.
 \label{Cw}
\ee
Finally, the joint distribution is 
\be 
 dg\, dw\, p(g,w,v=0) = \frac{1}{\sqrt{2\pi\sigma_v^2}}\,
 \frac{dg}{\sigma_0}\, \frac{e^{-g^2/2\sigma_0^2(1-\gamma_{gv}^2)}}{\sqrt{2\pi\,(1 - \gamma_{gv}^2)}}\ 
 dw\,\frac{e^{-(w-\langle w|v=0,g\rangle)^2/2\Sigma_{w|vg}^2}}{\sqrt{2\pi\,\Sigma^2_{w|vg}}}
 \label{pjoint}
\ee
with
\be
 \langle w|v=0,g\rangle = \sigma_w\,
 \frac{\gamma_{wg} - \gamma_{wv}\gamma_{vg}}{1 - \gamma_{gv}^2}\,\frac{g}{\sigma_0}
\ee
and
\be
 \Sigma^2_{w|gv} = \sigma_w^2\,
 \frac{1 - \gamma_{gv}^2 - \gamma_{wv}^2 -\gamma_{wg}^2 + 2\gamma_{gv}\gamma_{wv} \gamma_{wg}}{1 - \gamma_{gv}^2}.
\ee

\subsection{Implementing the constraints: Density fraction}

At matter-radiation equality, the density fraction in PBHs is 
\be
 \beta_\bullet 
  \equiv \frac{1}{\rho_{\rm eq}}\int dM_\bullet\, \frac{dn}{dM_\bullet}\,M_\bullet\,
  \equiv \int_{R_{\rm min}}^{R_{\rm max}} \frac{dR}{R}\,\frac{d\beta_\bullet}{d\ln R}
\ee
where $\rho_{\rm eq} = 3H_{\rm eq}^2$ is the radiation density evaluated at equality, $dn/dM_\bullet$ was given above, and the final expression defines $d\beta_\bullet/d\ln R$, the PBH fraction associated with comoving scale $R$ (approximately the PBH fraction which was created when the comoving horizon scale was $R$).  Explicitly, for the abundance associated with maxima of $g$,
 \be\label{beta}
 \beta_\bullet^{\rm max} =  \frac{a_{\rm eq}^3}{6(6\pi)^{3/2}}\int_{R_{\rm min}}^{R_{\rm max}} \frac{dR}{R}\left(\frac{R_{\rm eq}\sigma_2}{R\sigma_1}\right)^3\frac{H_{\rm eq}}{H_R}
 \int_0^\infty dw\, w\, \int_{g_{\rm min}}^{4/3} dg\, f(\frac{w+2g}{\sigma_\chi})\,p(g,w,v=0)\,{\cal K}\,\Bigl[\C(g) - \C_c(w)\Bigr]^{0.36},
 \ee
 where $g_{\rm min} = (4/3)\left(1-\sqrt{1- 3\C_c(w)/2}\right)$
 and we have used the fact that $H_{\rm eq} = (a_{\rm eq}R_{\rm eq})^{-1}$. The factor $a_{\rm eq}^3$ reflects the fact that the PBH density evolves as matter rather than radiation \footnote{If we were comparing $H_R$ with the Hubble scale at the beginning of the radiation era ($H_{\rm rad}$) rather than at the end ($H_{\rm eq}$), we would have had a single factor ($a_{\rm eq}$) rather than cubic, reflecting the slower evolution compared to radiation.  This is what is typically quoted in the literature.}.  

Similarly, the abundance of minima (in $g$) associated with $g>4/3$ is 
 \be\label{betam}
\beta_\bullet^{\rm min}=  \frac{a_{\rm eq}^3}{6(6\pi)^{3/2}}\int_{R_{\rm min}}^{R_{\rm max}} \frac{dR}{R}\left(\frac{R_{\rm eq}\sigma_2}{R\sigma_1}\right)^3\frac{H_{\rm eq}}{H_R}
 \int_{-\infty}^{0} dw\, |w|\times\cr
 \times \int_{4/3}^{g_{\rm max}} dg\, f(\frac{|w+2g|}{\sigma_\chi})\,p(g,w,v=0)\,{\cal K}\,\Bigl[\C(g) - \C_c(w)\Bigr]^{0.36}\theta\left(|w|-2g\right)\ ,
\ee
where $g_{\rm max}= (4/3)\left(1+\sqrt{1 - 3\C_c(w)/2}\right)$, and the constraint on $w$ comes from requiring the curvature to satisfy $w+2g<0$.  Since we know $g>4/3$, we want in particular $w<-8/3$ (for maxima we want $w+2g > 0$, which is always satisfied, since $g>0$ and $w>0$).

\section{The high peak limit}
For the $\beta_\bullet^{max}$ integral, it is useful to define the new variable
\be
 \bar g \equiv\frac{3}{4}\frac{g-\frac{4}{3}}{\sqrt{1-\frac{3}{2}\C_c(w)}},
 \label{gbar}
 \ee
since then 
\be
 \C(g) - \C_c(w)=\frac{2}{3}\left(1-\frac{3}{2}\C_c(w)\right)\left(1-\bar g^2\right)\ .
\ee
So, by using ${\cal K}\simeq 6$ and $R_{\rm max}\simeq R_{\rm eq}$ and making explicit the Hubble constant in terms of $R$, we get
\be\label{beta2}
 \beta_\bullet^{max}\simeq \frac{4}{3}\frac{a_{\rm eq}^3}{(6\pi)^{3/2}}
 \int_{R_{\rm min}}^{R_{\rm eq}} \frac{dR}{R}\frac{R_{\rm eq}}{R}
 \left(\frac{\sigma_2}{\sigma_1}\right)^3
 \int_{-1}^{0} d\bar g\, \Bigl[\frac{2}{3}\left(1-\bar g^2\right)\Bigr]^{0.36}
 \int_0^\infty dw\,w\left(1-\frac{3}{2}\C_c(w)\right)^{0.86}\,
  f(\frac{w+2\mathfrak{g}}{\sigma_\chi})\,p(g=\mathfrak{g},w,v=0)\ .\nonumber
\ee
where we have defined:
\be
\mathfrak{g}=\frac{4}{3}\left(1+\bar g\, \sqrt{1-\frac{3}{2}\C_c(w)}\right)\ .
\ee
If the correlations come from inflation, we generically expect all the $\sigma_j\ll 1$.\footnote{Note that however this might not be true for a very broad spectrum, c.f. Appendix \ref{C}.} We call this limiting case the {\it high peak limit}. Here, $p(w|g,v=0)$ is sharply peaked around its mean value $\bar w=\langle w|g,v=0\rangle$.  This defines the mean curvature profile needed to estimate  $M_\bullet$. Whether $\bar w$ is positive or negative depends on the power spectrum. If positive, $\beta^{max}_\bullet$ will dominate the abundance, otherwise, as long as $\bar w\leq -8/3$, $\beta_{\bullet}^{min}$ dominates.  In the high peak limit we then have
\be\label{beta30}
 \beta_\bullet^{min(max)}&\simeq& \frac{4}{3}\frac{a_{\rm eq}^3}{(6\pi)^{3/2}}
 \int_{R_{\rm min}}^{R_{\rm eq}} \frac{dR}{R}\frac{R_{\rm eq}}{R}
 \left(\frac{\sigma_2}{\sigma_1}\right)^3\frac{1}{\sqrt{2\pi}\sigma_v}\, \theta\left(\frac{\sigma_w}{\sigma_0} \frac{|
 \gamma_{wg} - \gamma_{wv}\gamma_{vg}|}{2 (1 - \gamma_{gv}^2)}-\epsilon\right)\,\times\cr
&\times&\int_{-1+\epsilon}^{\epsilon}d\bar g \Bigl[\frac{2}{3}\left(1-\bar g^2\right)\Bigr]^{0.36}\bar w\left(1-\frac{3}{2}\C_c(\bar w)\right)^{0.86}\, \left(\frac{\bar w+2\mathfrak{g}}{\sigma_\chi}\right)^3\, \frac{e^{-\mathfrak{g}^2/2\sigma_{g|v}^2}}{\sqrt{2\pi}\sigma_{g|v}}\ .
\ee
where we remind the reader that  $\bar w=\bar w(\bar g)$ and $\epsilon=0$ for the maximum and $1$ for the minimum. In \eqref{beta30}, we have introduced the notation $\sigma_{g|v}\equiv\sigma_g\sqrt{1-\gamma_{gv}^2}$, see also Appendix \ref{condpk}.  

We are now left with the integral in $\bar g$ which will depend on the shape of the power spectrum. However, to gain intuition, we can use the mean value theorem for integrals to obtain 
\be\label{beta3}
 \beta_\bullet^{min(max)}&\simeq& \frac{4}{3}\frac{a_{\rm eq}^3}{(6\pi)^{3/2}}
 \int_{R_{\rm min}}^{R_{\rm eq}} \frac{dR}{R}\frac{R_{\rm eq}}{R}
 \left(\frac{\sigma_2}{\sigma_1}\right)^3\frac{1}{\sqrt{2\pi}\sigma_v}\, \theta\left(\frac{\sigma_w}{\sigma_0} \frac{|
 \gamma_{wg} - \gamma_{wv}\gamma_{vg}|}{2 (1 - \gamma_{gv}^2)}-\epsilon\right)\,\times\cr
&\times& \Bigl[\frac{2}{3}\left(1-\bar g_s^2\right)\Bigr]^{0.36}\bar w_s\left(1-\frac{3}{2}\C_c(\bar w_s)\right)^{0.86}\, \left(\frac{\bar w_s+2\mathfrak{g}_s}{\sigma_\chi}\right)^3\, \frac{e^{-\mathfrak{g}_s^2/2\sigma_{g|v}^2}}{\sqrt{2\pi}\sigma_{g|v}}\ ,
\ee
where 
\be
\mathfrak{g}_s=\frac{4}{3}\left(1+\bar g_s\, \sqrt{1-\frac{3}{2}\C_c(\bar w_s)}\right)
\ee
and $|\bar g_s|<1$, or more specifically $-1<\bar g_s<0$ for the maximum and $0<\bar g_s<1$ for the minimum.  Since $\bar w_s\propto \mathfrak{g}_s$, for a given smoothing scale $R$,
\be
 \frac{d\beta_\bullet}{d\ln R}\sim e^{-\nu_{nl}^2/2}\ ,
\ee
where
\be
 \nu_{nl}^{min(max)}\equiv\frac{\frac{4}{3}\left(1+\bar g_s\sqrt{1-\frac{3}{2}\C_c(\bar w_s)}\right)}{\sigma_0\sqrt{1-\gamma_{gv}^2}}\ .
\ee
This shows that although the abundance is related to the mean profile (through $\bar w$), it is not dominated, in the exponential, by the threshold value related to that profile, which would imply $|\bar g_e|=1$. This is because, at the threshold $4/3$, the abundance is exactly zero. However, the exponential decay implies that $|\bar g_s|$ cannot be very different from $\sim 1$ \cite[e.g.][]{meilia}. Thus, we will further approximate
\be
\nu_{nl}^{min(max)}\simeq\frac{\frac{4}{3}\left(1\pm\sqrt{1-\frac{3}{2}\C_c(\bar w_s)}\right)}{\sigma_0\sqrt{1-\gamma_{gv}^2}}\equiv \frac{\delta_{th}}{\sigma_0\sqrt{1-\gamma_{gv}^2}}\ ,
\ee
where $\delta_{th}$ denotes the threshold associated with the mean profile $\bar w$.

\section{Comparison to previous work}

One of the main differences between our approach and what has been done to date, e.g. \cite{meilia, jaume, vicente, chris, riotto}, is that we calculate the abundance of PBHs related to the all statistical realisations of over-density profiles and smoothing scales, rather than focusing only on the mean profile at a given smoothing scale.  Nevertheless, as we have seen, in the high peak limit the dominant contribution to the abundance is given by the mean profiles per given smoothing scale $R$. Fixing the smoothing scale, we can then roughly compare our non-linear statistics in the high peak limit with previous other approaches. 

\subsection{Linear approach}
The PBH abundance calculated by the use of the linearized over-density \cite{meilia} and related to the scale $R$ correspondent to the maximum of the compaction function constructed on the mean over-density, was found to be proportional to $e^{-\nu_l^2/2}$ where \cite{vicente}
\be
\nu_{l}\simeq\frac{3}{4}\delta_c\frac{\sqrt{\int \frac{dk}{k} (kR)^4{\cal P}(k)}}{\int \frac{dk}{k} (k R)^4\, {\rm sinc}(k R)\, {\cal P}(k)}\ .
\ee
Assuming the non linear profile to be the same as the linear one, $\delta_c$ was taken to coincide with the full non-linear threshold $\C_c$ calculated by considering $\frac{\delta\rho(r,t)}{\delta\rho(0,t)}\simeq\frac{\Delta(r,t)}{\Delta(0,t)}$. In other words it was assumed that all non-linearities were encoded in the amplitude rather than in the shape. At least at second order expansion in the curvature $\zeta$, $\frac{\delta\rho(r,t)}{\delta\rho(0,t)}\simeq\frac{\Delta(r,t)}{\Delta(0,t)}$ seems to be a good approximation \cite{riotto}. However, this approach suffers from an ambiguity. Because the constraint on the maximum of the compaction function at $R$ is not implemented, 
\be
\delta_{c}=\frac{4}{3}\left(1\pm\sqrt{1-\frac{3}{2}\delta_{l}}\right)\ ,
\ee
where $\delta_l$ is the threshold related to the mean profile of $\Delta$. Since this approach treats both thresholds equally, in principle the total abundance should be the sum of the two, however, the exponential suppression would select
\be
\nu_{l}\simeq\left(1-\sqrt{1-\frac{3}{2}\delta_l}\right)\frac{\sqrt{\int \frac{dk}{k} (kR)^4{\cal P}(k)}}{\int \frac{dk}{k} (k R)^4\, {\rm sinc}(k R)\, {\cal P}(k)}\ .
\ee

\subsection{Curvature perturbation approach}

To overcome the difficulty of dealing with the non-linear relation between the curvature perturbation $\zeta$ and the over-density $\delta\rho/\rho$, the authors of \cite{jaume} have considered the statistics of high peaks of $\zeta$ and argued that their abundance should generically match those of $\delta\rho/\rho$. This leads to a factor proportional to $e^{-\nu_\zeta^2/2}$ where
\be
\nu_\zeta=3\left(1\pm\sqrt{1-\frac{3}{2}\delta_\zeta}\right)\frac{\sqrt{\int \frac{dk}{k} {\cal P}(k)}}{\int \frac{dk}{k} (k R)^2W_{TH}(k R){\cal P}(k)}\ ,
\ee
 and $R$ corresponds now to the maximum of the compaction function associated to the mean $\zeta$ profile with $\delta_{\zeta}$ its threshold. This value was assumed to be $\delta_{\zeta}=0.5$. With our fitting formula \eqref{deltac} we would not need to guess it. As before this approach suffer from the ambiguity of the threshold, so we can again approximate 
 \be
 \nu_\zeta\simeq 3\left(1-\sqrt{1-\frac{3}{2}\delta_\zeta}\right)\frac{\sqrt{\int \frac{dk}{k} {\cal P}(k)}}{\int \frac{dk}{k} (k R)^2W_{TH}(k R){\cal P}(k)}\ .
 \ee
 Unfortunately, this approach has a serious drawback due to the IR divergences that must be regulated. This has been circumvented in \cite{jaume} by only considering power spectra that go to zero rapidly as $k\rightarrow 0$. Although in real situations some regularization must be done, here, we will only consider such ``regularized'' spectra.
 
\subsection{Comparison I: ${\cal P}(k)$ with a narrow feature in $k$}

 Both linear and curvature perturbation approaches miss the factor $\sqrt{1-\gamma_{gv}^2}$. This factor arises from the constraint that the compaction function must be a maximum on scale $R$, a condition that is not enforced in either \cite{meilia} or \cite{jaume}.  Since this term appears in an exponential, previous approaches can only really approximate ours in the limit where $\sigma_{g|v}\rightarrow \sigma_{g}$.  
 
 Generically $\C_c(\bar w_s)\neq \delta_l\neq \delta_\zeta$ as each of them corresponds to a different profile. It is however tempting to consider
 $\C_c(\bar w_s)\sim \delta_l\sim \delta_\zeta$. This would suggest that the linear threshold is $3$ times smaller than the non-linear one, or the one obtained by the use of $\zeta$. Nevertheless, as it has been numerically shown in \cite{riotto}
 \be
 \left(1-\sqrt{1-\frac{3}{2}\delta_\zeta}\right)<\left(1-\sqrt{1-\frac{3}{2}\delta_l}\right)\ .
 \ee
 Thus, at least for this case, the assumption $\delta_l= \delta_\zeta$ turns out to be inconsistent, while $\C_c\sim\delta_\zeta$ will not be a bad approximation.
 
Apart from the correlation taking into account the constraint on the maximum of the compaction function, we also see discrepancies in the other correlators.  These might be minimized for very peaked profiles (and maximized the other way around).  Indeed, considering 
 \be\label{power}
 {\cal P}\sim k_p {\cal P}_0 \delta(k-k_p)\ ,
 \ee
we see that, for the curvature perturbations approach,  
\be
\frac{\sqrt{\int \frac{dk}{k} {\cal P}(k)}}{\int \frac{dk}{k} (k R)^2W_{TH}(k R){\cal P}(k)}\sim \frac{1}{(k_p R)^2 W_{TH}(k_p R)\sqrt{{\cal P}_0}}\sim \sigma_g^{-1}\ .
\ee 
In contrast, for the linear approach,
\be
\frac{\sqrt{\int \frac{dk}{k} (kR)^4{\cal P}(k)}}{\int \frac{dk}{k} (k R)^4\, {\rm sinc}(k R)\, {\cal P}(k)}\sim \frac{W_{TH}(k_p R)}{{\rm sinc}(k_p R)}\sigma_g^{-1}\ .
\ee
However, because ${\rm sinc}(k_p R)<W_{TH}(k_p R)$ for the values of $k_p R$ of most interest, and because the non-linear threshold is smaller than the linear one, we can roughly approximate, for the peaked profile,
\be
\sqrt{1-\gamma_{gv}^2}\,\nu_{nl}\sim\nu_{\zeta}\sim\nu_{l}\ ,
\ee
where we remind the reader that we have assumed that the three approaches have the same overall numerical factors and R. This assumption seems not to be so bad though \cite{riotto}, at least for the case of a very narrow power spectrum. 

In this promising case, we indeed are going to see that all statistics coincide. For a power spectrum of the form \eqref{power}, $\Sigma_{w|vg}\rightarrow 0$ and so the integral in $w$ becomes a delta function centered in $w=\bar w$. This singles out only the mean profile (see also \cite{jaume2}). In addition, $\gamma_{gv}^2\rightarrow 1$ for $R$ except for one special value.\footnote{CG would like to thank Jaume Garriga for pointing this out.}  When $\gamma_{gv}\rightarrow 1$, $p(g|v=0)$ is a delta function centered on $g=0$. This is too low to produce PBHs, so such positions have $\beta_\bullet\rightarrow 0$.

\begin{figure}
  \centering
  \includegraphics[width=0.8\hsize]{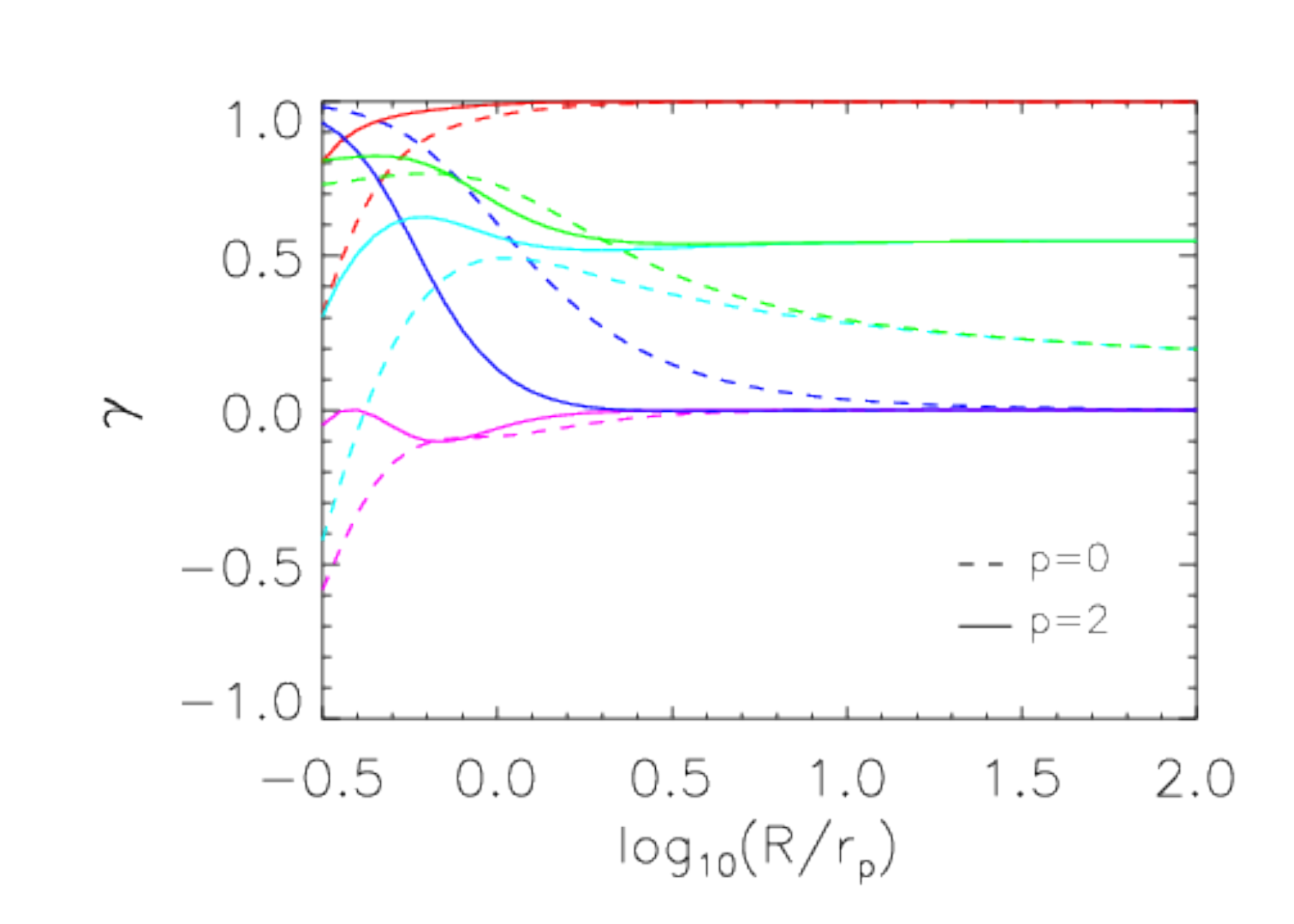}
  \caption{Correlation coefficients associated with the model in which ${\cal P}(k) = A_s\, (kr_p)^p \exp(-kr_p)$.  Blue, cyan, and green show $\gamma_{gv}, \gamma_{gw}$ and $\gamma_{g\chi}$, whereas red and magenta show $\gamma_{w\chi}$ and $\gamma_{wv}$. The limit $R/r_p\rightarrow 0$ corresponds to a highly peaked profile. There, we see a numerical confirmation of our analysis for the delta function power spectrum. In the $R/r_p\gg 1$ case we instead have a modeling of a broad power spectrum. Here, all the Pearson coefficients become approximately constant.}
  \label{gammaPBH}
\end{figure}

The Pearson correlation coefficients are  $\gamma^2_{gv}\propto\frac{1}{\sigma_g^2\sigma_v^2}$. While $\sigma_g^2>0$, the correlator describing the peak of $g$ in $\vec{x}_0$ ($\sigma_v$) has a zero at $R=R_c \simeq 2.74\times k_p^{-1}$. Thus, at that point, care has to be taken in implementing the delta function in the power spectrum. Taking for example a succession of symmetric rectangles with base $2\epsilon$ around the peak $k_p$ and heights $1/(2\epsilon)$, one finds that $\gamma_{gv}$ has the following structure
\be
\gamma_{gv}=\frac{\tilde\sigma_v+\epsilon^2\,\tilde\sigma_v\, n(R)+{\cal O}(\epsilon^3)}{|\tilde\sigma_v|+\epsilon^2\, m(R)+{\cal O}(\epsilon^3)}\ ,
\ee 
where $m(R)$ and $n(R)$ are non vanishing functions and $\tilde\sigma_v\equiv\lim_{\epsilon\rightarrow 0}\sigma_v$. We then immediately see that if $R\neq R_c$, in the limit $\epsilon\rightarrow 0$ which corresponds to the Dirac delta function, $\gamma_{gv}\rightarrow 1$, while in $R=R_c$, $\gamma_{gv}\rightarrow 0$. The same as for $\gamma_{gv}$ happens for $\gamma_{wv}$ while $\gamma_{wg}^2\rightarrow 1$ so that $\Sigma_{w|gv}\rightarrow 0$ (see also Fig. \ref{gammaPBH} for a numerical confirmation of this analysis). Thus, as already discussed, even for $R=R_c$ the distribution of profiles is distributionally peaked at $w=\bar w$.

The value $R=R_c$ corresponds also to the radius at which the compaction function constructed on the mean linear density profile \cite{meilia} and/or the mean curvature profile \cite{jaume} becomes maximal. At this radius $\frac{\bar w}{g}\simeq 5.53$. In this case then the assumption that the dominant contribution to the abundance is related to a maximum of $g$ is also correct. The value $\frac{\bar w}{g}\simeq 5.53$ corresponds to $\delta_{th}\simeq 0.49$, which is very similar to what found in previous literature \cite{meilia}, or assumed in \cite{jaume}. Interestingly also, for $R=R_c$ $W_{TH}(k_p R_c)/{\rm sinc}(k_p R_c)\simeq 3$ taking care of the threshold difference from the linear to the non-linear case, in agreement to what we have discussed above.

Finally, at this radius, $\gamma_{gv}\rightarrow 0$ and thus
\be
\nu_{nl}\Big|_{R=R_c}\simeq \nu_{l}\Big|_{R=R_c}\simeq \nu_{\zeta}\Big|_{R=R_c}\ .
\ee 
One would then be worried that, because we actually have an integral in $R$, the abundance of an isolated point would be zero. However, the integrand in $R$ is also proportional to $\sigma_v^{-1}$ effectively generating a delta function at $R=R_c$. This also implies that, for a very peaked ${\cal P}$, the mass spectrum is also sharply peaked.

We want to end this section though by warning the reader that, although the exponential suppression of the three approaches are similar, the exact abundance might differ by several orders of magnitude due to the exact details of $\beta_\bullet$, even in the high peak limit eq. \eqref{beta3}.

\subsection{Comparison II:  Excess power over broader range of $k$}
To model a power spectrum which contributes over a broad range of $k$ we consider ${\cal P} = A_s\, (kr_p)^{p}\,\exp(-kr_p)$ for some $r_p$ smaller than any $R$ of interest.  For $p>0$, ${\cal P}\to 0$ at small $k$, insuring that this modification will not affect the CMB.  In addition, it is expected that $p<5$ \cite{chris2}.  
For integer $p$, all the correlators of interest (Eq.~\ref{gammas}) can be computed analytically; we provide explicit expressions in Appendix~\ref{C}.  This shows that, in the  $\rho\equiv R/r_p\gg 1$ limit, $\sigma_j^2\propto \rho^{2j}$ for all $p>0$. In the $R/r_p\gg 1$ limit, all the $\gamma$s tend to constant values. In particular, $\gamma_{w\chi}\to 1$, $\gamma_{gw}\to\gamma_{g\chi}$, $\gamma_{wv}\to 0$ and $\gamma_{gv}\to 0$.  That $\gamma_{gv}\to 0$ is significant, since then $\nu_{g|v}\to \nu_g$: the constraint on $v$ does not matter for the distribution of $g$.

Figure~\ref{gammaPBH} shows the correlation coefficients associated with this model.  Blue, cyan, and green show $\gamma_{gv}, \gamma_{gw}$ and $\gamma_{g\chi}$, whereas red and magenta show $\gamma_{w\chi}$ and $\gamma_{wv}$.  Note that all these $\gamma$s tend to constant values at large $R/r_p$.  In particular, at $R/r_p\gg 1$, $\gamma_{w\chi}\to 1$, $\gamma_{wv}\to 0$ and $\gamma_{gv}\to 0$.  

We are particularly interested in the combination $\gamma_{wg} - \gamma_{wv}\gamma_{vg}$, since this controls the sign of $\bar w$.  For $0.49\leq g<4/3$ we require $w>0$, so if the sign of $\bar w$ is negative, $w>0$ will be extremely unlikely.  Figure~\ref{gammaPBH} shows that $\gamma_{wv}\gamma_{vg}\ll \gamma_{wg}$ for all $R/r_p>3$, and, since $\gamma_{wg}$ is positive, $\bar w$ is also.  This means that PBH abundances will be dominated by the peaks associated with $0.49\leq g<4/3$; the contribution from $g>4/3$ which we argued require $w<0$ will be suppressed.   However, for $R/r_p<1$ or so, both $\gamma_{gw}$ and $\gamma_{\chi w}$ change sign.  As a result, for sufficiently small $R/r_p$, $\bar w$ can be negative; in this case, it is the contribution from $0.49\leq g<4/3$ that is suppressed.

The case of large smoothing radius requires an additional comment. Because the $\sigma_j$s grow with $\rho$, the high peak approximation is bound to fail at some $R\gg r_p$ (c.f. Appendix \ref{C}). The reason is that at some point all the $\sigma_j$s (for $j>0$) become larger than one.  In this case the rare peak approximation, on which our simplified statistics is based, would also fail.  However, it is easy to see, c.f. Appendix \ref{C}, that
at larger $R$, $\beta_\bullet$ stabilises to a constant. This can also be appreciated by looking at Fig. \ref{fig:betaR}, which shows how the sum of Eqs. (\ref{beta}, \ref{betam}) -- evaluated numerically using the full joint distribution of $g$ and $w$ (Eq.~\ref{pjoint}) -- grows as the upper limit in $R$ is increased. 

Finally, we have also explored another broad ${\cal P}$ parametrization:  we use a flat power spectrum ranging from $k=k_{min}$ to $k=k_{max}$ with $k_{min}\ll k_{max}$, as the one used in \cite{meilia} to model the broad case, and taking $\delta_l(R_b)\simeq\C_c(R_b)$ \cite{riotto}, where $R=R_b$ is the radius associated to the maximum of the compaction function calculated with the mean over-density profile. We then find, in the high peak limit where $k_{max}R$ is not too large, that $\nu_{nl}(R_b)\simeq0.4\times\nu_{l}(R_b)$. In other words, the linear analysis over-estimates the PBH abundance. This is because $g$ is unbounded in the linear analysis, so the configuration space which gives rise to PBHs is larger.

On the other hand, the estimate based on curvature statistics greatly under-estimates the PBH abundance due to the IR divergences. So in general we have that, for a broad power spectrum,
\be
\nu_\zeta\gg \nu_{nl}> \nu_{l}\ .
\ee

\subsection{Comparison III: PBH constraints on ${\cal P}(k)$}

\begin{figure}
  \centering
  \includegraphics[width=0.45\hsize]{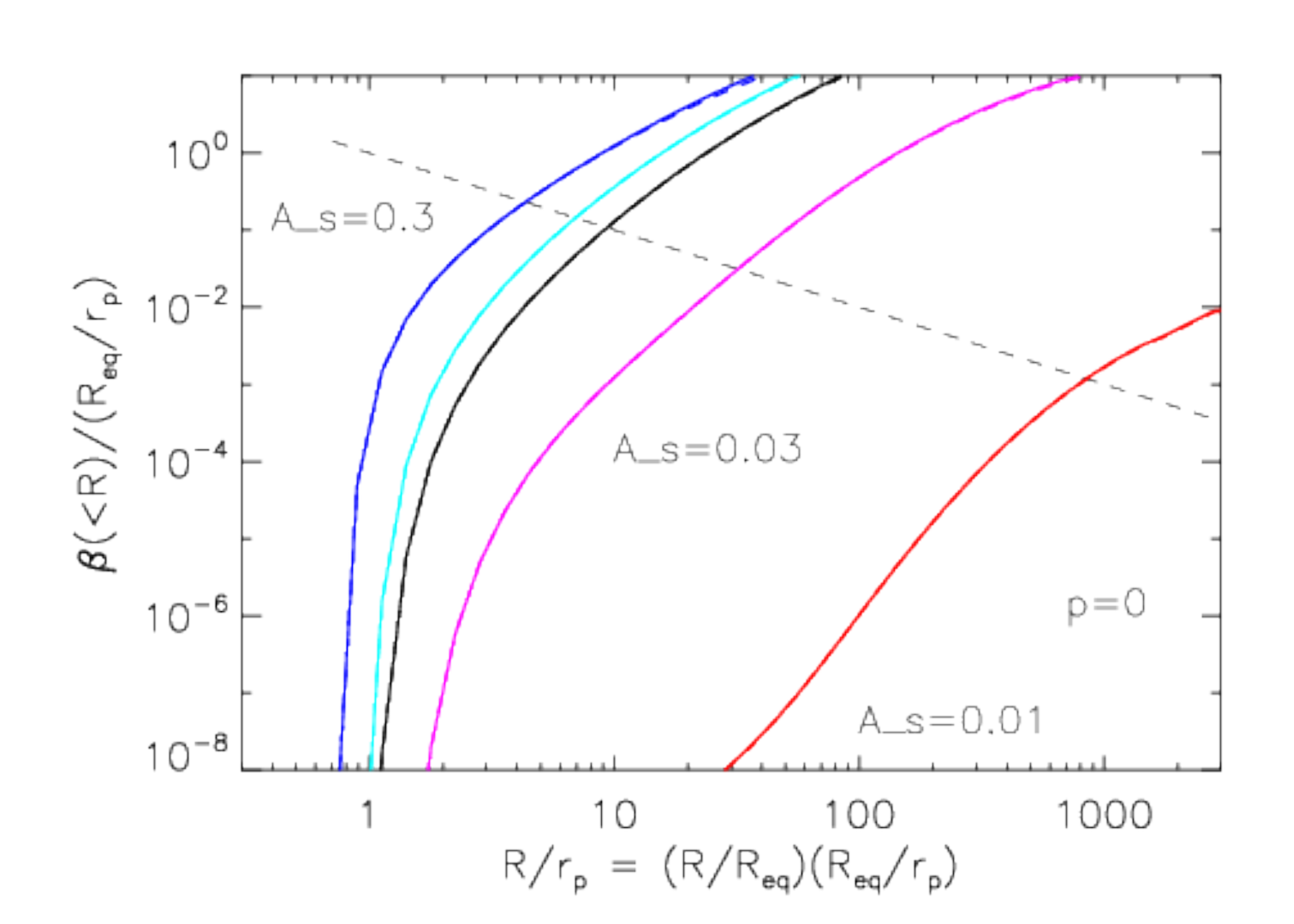}
  \includegraphics[width=0.45\hsize]{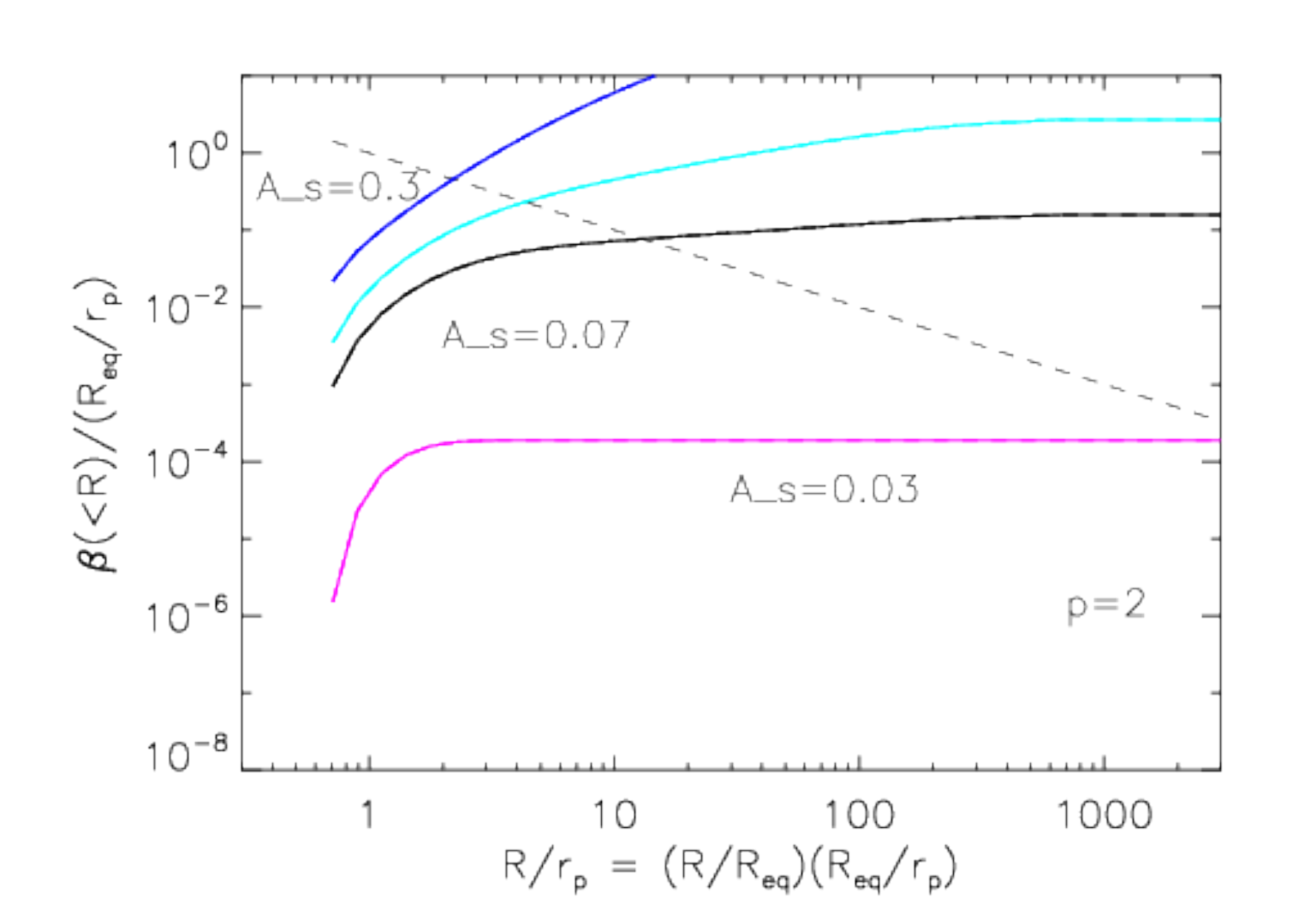}
  \caption{Increase of mass density in PBHs as $R$ increases, for a number of choices of the amplitude $A_s$ of the power spectrum ${\cal P}(k)=A_s (kr_p)^p\,exp(-kr_p)$ with $p=0$ (left) and $p=2$ (right).  Dashed line shows the limit in which PBHs account for all the matter density.  \label{fig:betaR}}
\end{figure}

The predicted density in PBHs depends on the shape and amplitude of ${\cal P}(k)$.  To illustrate this, Figure~\ref{fig:betaR} shows how the cumulative density in PBHs increases with $R$ for $A_s = [0.01,0.03,0.07,0.1,0.3]$ (bottom to top).  The left and right hand panels show results for the shape parameter $p=0$ and $p=2$ (the lowest value of $A_s$ is not visible for this case).  For the family of ${\cal P}(k)$ we are considering, the predictions also depend on the scale $r_p$ beyond which the power drops exponentially.  Since $\beta_\bullet$ of Eqs.~(\ref{beta}) and~(\ref{betam}) express the PBH density in units of $\rho_{\rm eq}$, it is convenient to consider how the predictions depend on the ratio $R_{\rm eq}/r_p$.

Eqs.~(\ref{beta}) and~\ref{betam}) show that, for a given $A_s$ and $p$, $\beta_\bullet/(R_{\rm eq}/r_p)$ is a function of $R/r_p = (R/R_{\rm eq})(R_{\rm eq}/r_p)$.  This explains why the axes of Figure~\ref{fig:betaR} have been scaled in this way.  Changing $R_{eq}/r_p$ shifts the curves up and left or down and right.  The straight dashed line shows the locus of $\beta_\bullet=1$ at $R/R_{\rm eq}=1$ as $R_{\rm eq}/r_p$ varies.  Values above or to the right of this line indicate that, for this choice of $A_s$ and $p$, the predicted PBH abundance is too large, as the entire density within the horizon at equality would be due to PBHs.  Clearly, for a given $p$, there is a degeneracy between $A_s$ and $r_p$.  This is in the expected sense:  decreasing $r_p$ increases $R_{\rm eq}/r_p$, meaning that more modes contribute to the power within $R_{\rm eq}$; hence, the overall amplitude must be reduced so as to not overproduce PBHs.  This provides a simple illustration of how PBH abundances constrain the primoridal ${\cal P}(k)$.  

\begin{figure}
  \centering
  \includegraphics[width=0.45\hsize]{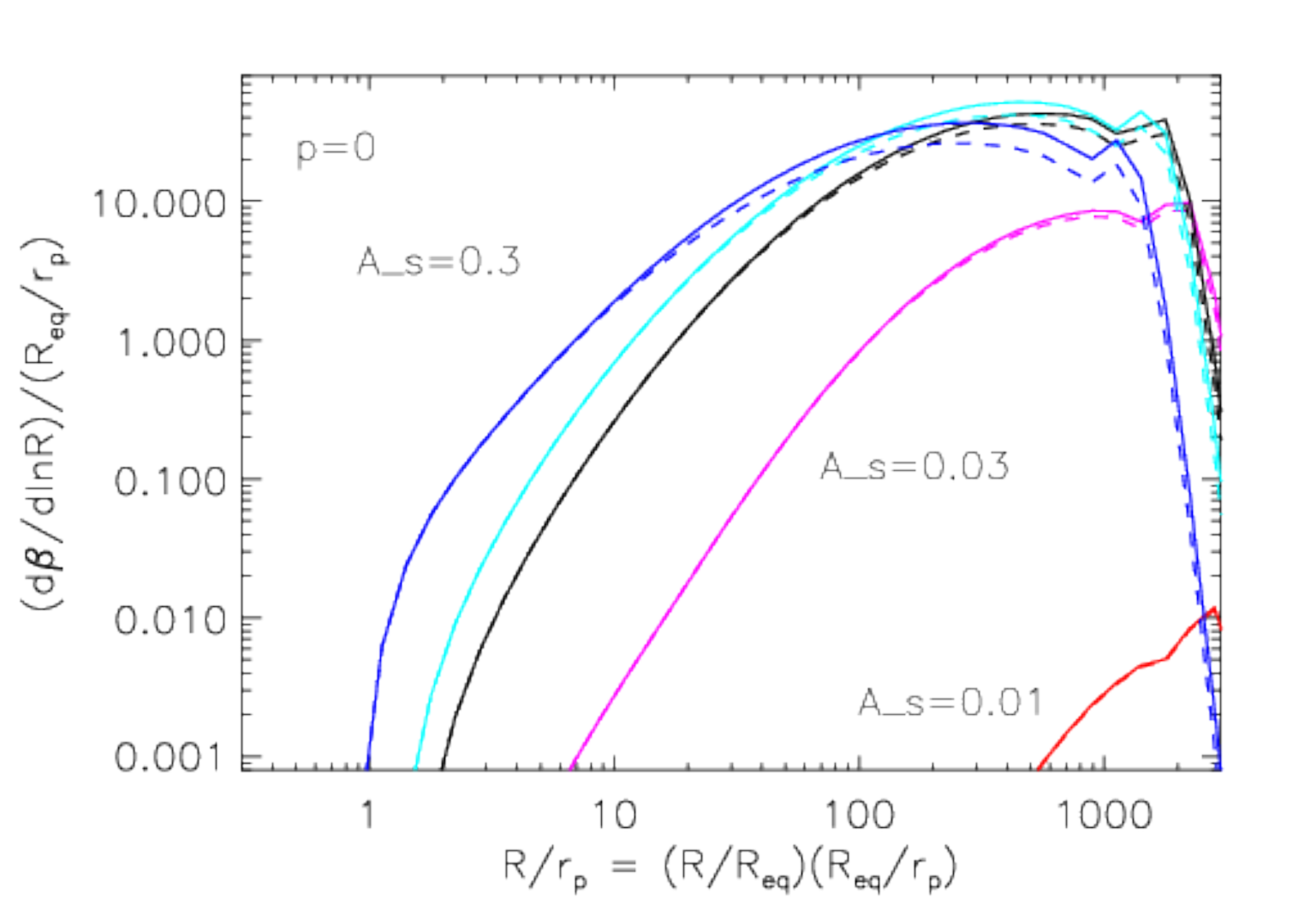}
  \includegraphics[width=0.45\hsize]{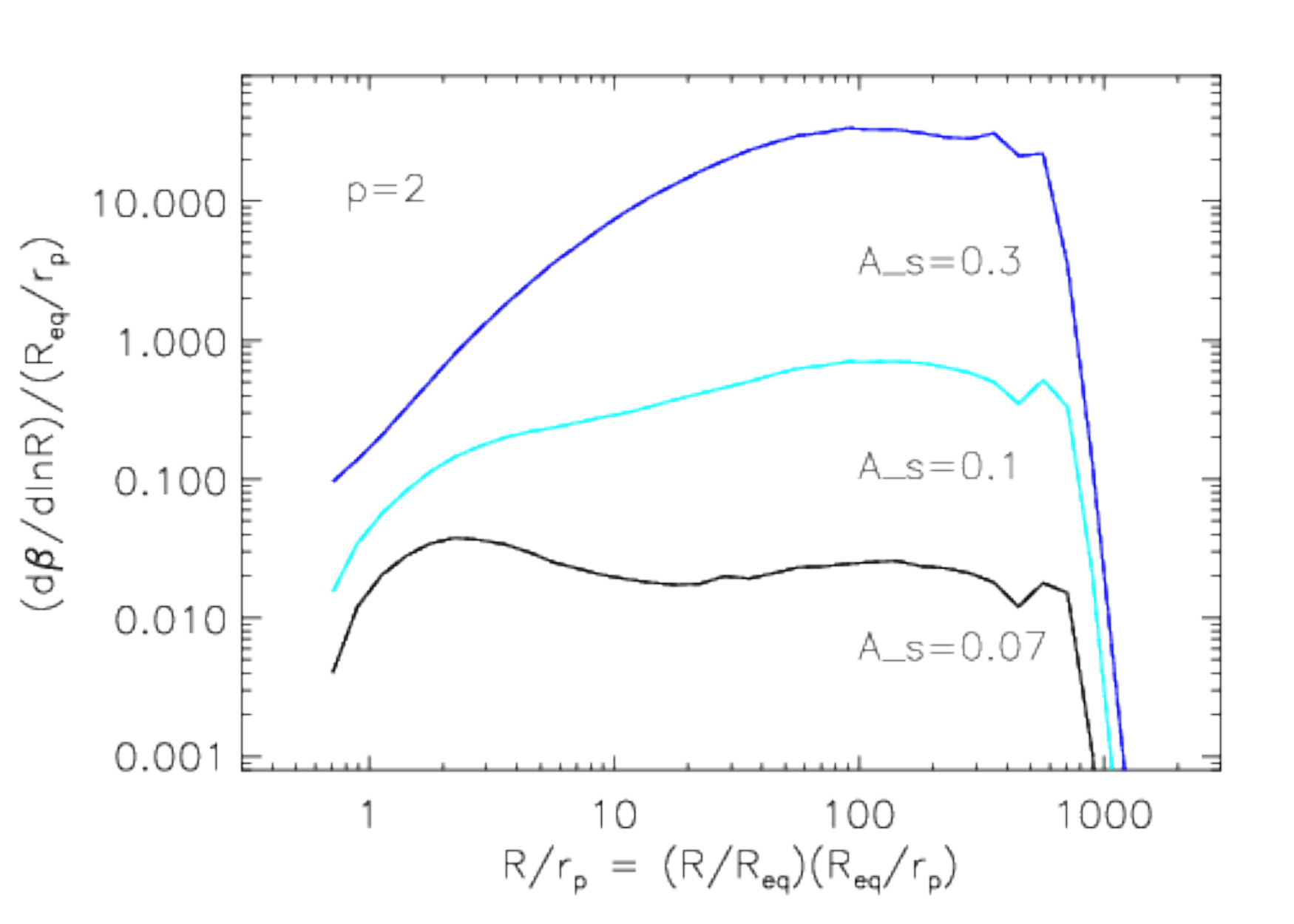}
  \caption{Dependence of $d\beta_\bullet/d\ln R$ on the amplitude and shape of the power spectrum with $p=0$ (left) and $p=2$ (right).  Dashed curves show the contribution from maxima, and solid curves add the contribution from minima as well.  Minima matter for $p=0$ but not for $p=2$.\label{fig:dbetadlnR}}
\end{figure}

Finally, because we made the point that minima in $g_R$ may make a non-negligible contribution to the counts, Figure~\ref{fig:dbetadlnR} shows the total differential counts $d\beta_\bullet/d\ln R$ (solid) and the contribution from maxima only (dashed).  Clearly, minima matter for $p=0$, but not for $p=2$.  This plot also illustrates that the range of $R$ (and hence times) over which PBHs form depends on $A_s$ and $p$.

\section{Conclusions}
Primordial Black Holes might account for the whole of the missing dark matter. If generated during the radiation dominated epoch of our Universe, they do not need to be very copious, on the contrary, because their density scales like matter, they must be extremely rare at formation. 

The typical mechanism for PBH formation during the radiation dominated era is the collapse of rare large inhomogeneities in the energy density produced during inflation at cosmological horizon re-entry (for alternative mechanisms see e.g. \cite{kusenko}). Because they are rare, they depend exponentially on the exact statistical distribution of the over-densities that would collapse into PBHs. So far, this statistical distribution has been estimated using the linear statistics (either in terms of linear over-density or in terms of curvature perturbations). In this paper, assuming a gaussianly distributed curvature perturbation, we instead provided for the first time the full non-linear statistics of PBH and their mass distribution. 

PBHs are formed whenever the ``gravitational potential'' at super-horizon scales, or more precisely the compaction function, at its maximum exceeds a certain critical threshold. The (fully non-linear) compaction function is a quadratic polynomial of the smoothed {\it linear} over-density so, for gaussian statistics of curvature perturbations, the statistics of the compaction function is a non-central chi-square. 

The difference between the linear and non-linear statistics goes beyond the mere choice of the correct statistical variable. In the linear statistics approaches PBHs are associated with peaks which exceed a certain threshold. This threshold was obtained by considering the shape of a mean profile, related to a given power spectrum of curvature perturbations, and plugged into the deterministic relation between the linear over-density and the compaction function \cite{meilia} (similarly in the curvature perturbation approach \cite{jaume}). This approach leaves the question of whether the threshold so calculated can be used for all possible statistical realizations of the over-densities that would collapse into PBHs. In our approach, the thresholds for each possible over-density realization are included self-consistently. 

An additional difference between our approach and previous work is the choice of the statistical conditions for PBH formation. In the linear analyses the conditions for a PBH are that a) there exist a maximum (peak) for the linear over-density, b) this maximum is larger than a certain threshold. In the non-linear statistics the conditions are quite different: a) there exist an extremum (either a maximum or a minimum) of the linear over-density, b) there exist a smoothing radius $R$ for which the compaction function is maximal and c) the compaction function is larger than a threshold that is related to the specific realization of the linear over-density.

Although the results based on non-linear statistics appear remarkably different from the linear-based ones, for an extremely peaked power spectrum, the order of magnitude of PBH abundance is very similar. In contrast, for a broad power spectrum, the results are quite different.  First of all, the mean over-density dominates the statistics only up to a moderately large smoothing scale. After that, the full statistical realisation of profiles must be taken into account. Secondly, even in the case of small smoothing scales, linear over-density statistics over-estimate the PBHs abundance while the linear curvature statistics under-estimate it relative to our nonlinear one. For the linear over-density approach, this over-estimation is because the amplitudes that would form a black hole seem to be only bounded from below. This is however an artifact of the linear analysis as, after a certain amplitude for the linear over-density, the compaction function (the non-linear gravitational potential) starts to decrease.  On the other hand, the curvature perturbation approach suffers from the fact that extremes of over-densities might not always coincide with extremes of curvature perturbations (while the contrary is always true). Thus, generically, the curvature perturbation approach, based on counting the extremes of curvature perturbations, under-estimates the PBH abundance.  
 
\section*{Note added}
Shortly after our results were presented at the focus week on primordial black holes at the Kavli IPMU by CG \cite{focus}, a very similar analysis to ours appeared on the arXiv \cite{suyama}.  This work only sketches the argument that leads to the first equation of our Appendix~\ref{x2p}, but does not proceed further.

\appendix

\section{From point process to statistics}\label{x2p}
Let $\vec{x}_\bullet$ denote a position around which a PBH forms in a single realization of the field $\g$.  The mean number density of such positions is given by adding up the point process of such PBH positions and dividing by the total volume.  Averaging this over all realizations of the field yields 
\be
\frac{dn}{d\ln R}
= \left\langle\sum_\bullet \delta_{\rm D}^{(3)}(\vec{x}-\vec{x}_\bullet)\right\rangle
= \left\langle|J|\,\delta_{\rm D}^{(3)}(\vec{\nabla} g_R)\,\vartheta(\lambda)\,\delta_{\rm D}(v_R) \,\vartheta(w_R)\,C_{\rm max}(g_R)\right\rangle
\label{fullstats}
\ee
where the first delta function expresses the requirement that the spatial gradient equal zero, the step function requires the second derivatives to have the right sign (this can be written as the requirement that the three eigenvalues of the matrix $\nabla^2_{ij}$ all have the right sign; it is conventional to use $\lambda$ to denote these eigenvalues, so we want the smallest of these $|\lambda|$ to have the right sign), the second delta function requires $g_R$ to be an extremum on smoothing scale $R$ and the final step function requires this extremum to be a maximum (rather than a minimum), and $C_{\rm max}(g_R)$ expresses the requirement that we are typically only interested in maxima for which $g_R$ lies between certain values (e.g. $0.49\le g_R\le 4/3$ of Eq.~\ref{conditions}).  Finally, $|J| = \det[\partial\{v_R,\vec{\nabla}g_R\}/\partial\{R,\vec{x}\}]$ is the Jacobian determinant (associated with the delta-functions) which transforms from spatial positions $\vec{x}$ and scales $R$ to variables $\vec{\nabla}g_R$ and $v_R$.  If we set $\eta_i\equiv \nabla_i g$, then $\partial\eta_i/\partial\ln R = \nabla_iv$ and 
\be
|J| = \begin{vmatrix} 
	\partial v/\partial\ln R & \nabla_x v & \nabla_y v& \nabla_z v  \\
	\nabla_x v & \nabla^2_{xx} g & \nabla^2_{xy} g & \nabla^2_{xz} g \\
	\nabla_y v & \nabla^2_{yx} g & \nabla^2_{yy} g & \nabla^2_{yz} g \\
	\nabla_z v & \nabla^2_{zx} g & \nabla^2_{zy} g & \nabla^2_{zz} g 
\end{vmatrix}
= \det({\vec{\nabla}\vec{\eta}})
\left(\frac{\partial v}{\partial\ln R} - \nabla_iv\Bigl(\nabla^2_{ij}g\Bigr)^{-1}\nabla_j v\right).
\ee
To appreciate the physical significance of the two terms, it is useful to study the case in which $\vec{\nabla} v=0$ (in addition to $\nabla_i g_R=0$ and $v_R=0$).  Then $|J|$ simplifies to $|\det({\vec{\nabla}\vec{\eta}})|\, |\partial v/\partial\ln R|$.  Since $\partial v/\partial\ln R = v_R-w_R$, and $v_R=0$, we have $|J| = |\det({\vec{\nabla}\vec{\eta}})|\, |-w_R|$: the first factor is the spatial peaks constraint of \cite{bbks} and the second is a result of requiring the profile to have $v_R=0$.  This simple factorization of the two requirements is what appears explicitly in the analysis in the main text (the $\det({\vec{\nabla}\vec{\eta}})$ term gives rise to the $f(\chi/\sigma_\chi)$ term).

Although generically $\vec{\nabla} v\ne 0$, for extremely rare peaks, i.e. the ones which matter most for PBHs, $\vec{\nabla} v\sim 0$ whenever $\vec{g}_R = 0$. Nevertheless, for completeness, we shall also discuss the case in which $\vec{\nabla} v\ne 0$.  Here, it is useful to consider the fact that if there is a peak at position $\xo$ when the smoothing scale is $R$, then a small change $R$ may result in a slightly shifted peak.  To impose the spatial peak constraint across smoothing scales (i.e. along the path traced by the shifting peak position as $R$ changes), we must require 
\be
\frac{d\eta_i}{dR} \equiv \frac{\partial \eta_i}{\partial R} + \frac{d\vec{x}_{{\rm pk}}}{dR}\cdot\vec{\nabla}\eta_i = 0
\qquad{\rm which\ implies}\qquad 
\frac{\partial \eta_i}{\partial\ln R}\Bigl(\nabla_i\eta_j\Bigr)^{-1} = -\frac{d\vec{x}_{j{\rm pk}}}{d\ln R}.
\ee
This means that 
\be 
|J| = \det({\vec{\nabla}\vec{\eta}})
\left(\frac{\partial v}{\partial\ln R}
+ \frac{d\vec{x}_{\rm pk}}{d\ln R}\cdot\vec{\nabla} v\right)
= \det({\vec{\nabla}\vec{\eta}}) \left|\frac{dv}{d\ln R}\right|,
\ee
which is again in factorized form for the spatial and scale curvatures, but now we see that the scale constraint is really in terms of the total derivative (with respect to $R$).  In particular, requiring $dv/d\ln R \le 0$ means 
\be
\frac{\partial v}{\partial\ln R} = v-w 
\le \nabla_iv\Bigl(\nabla^2_{ij}g\Bigr)^{-1}\nabla_j v.
\ee
Since we are only interested in peaks with $v=0$, this constrains $w$.  Clearly, if $\nabla_i v = 0$ then this requires $w\ge 0$, whereas if $\nabla_i v \ne 0$ 
then the lower limit on $w$ is increased to $-\nabla_iv\Bigl(\nabla^2_{ij}g\Bigr)^{-1}\nabla_j v$ (i.e. at a peak in $g$, $\nabla_iv\Bigl(\nabla^2_{ij}g\Bigr)^{-1}\nabla_j v\le 0$, so the profile must fall more steeply from its maximum value as $R$ changes).  In the main text, increasing the lower limit on $w$ will have the result of making most PBHs have $\C_c$ closer to the maximal value of $2/3$.  Alternatively, since $w = \chi - 2g$, 
\be
\chi \ge 2g - \nabla_iv\Bigl(\nabla^2_{ij}g\Bigr)^{-1}\nabla_j v.
\ee
This shows that $\nabla_i v\ne 0$ implies larger $\chi$ (i.e. larger peak curvature).  

We now consider the conditional distribution of $\nabla_i v$ given $\eta_i\equiv\nabla_i g = 0$:  $p(\nabla_i v|\eta_i=0)$.  This will be gaussian, with mean zero (since it is proportional to the constraint $\eta_i=0$) and variance
\be
\Sigma^2_{\eta_v|\eta} \equiv \langle(\nabla_i v)^2\rangle\,(1 - \gamma_{\eta_v\eta}^2), \quad{\rm where}\quad 
\gamma_{\eta_v\eta}^2 = \frac{\langle \eta_i\nabla_iv\rangle^2}{\langle\eta_i^2\rangle\langle(\nabla_i v)^2\rangle}\quad {\rm with}\quad
\langle \eta_i\nabla_iv\rangle
= \frac{1}{2}\frac{\partial\langle \eta_i^2\rangle}{\partial\ln R}
= \frac{1}{2}\frac{\partial\,(\sigma_1^2/3)}{\partial\ln R}.
\ee
Note that $\vec{\nabla} v$ does not correlate with any of the other statistical variables.  This makes it relatively easy to incorporate into the analysis in the main text.  The crude intuition is that integrating over the $p(\nabla_i v|\eta_i=0)$ will replace $\nabla_iv\Bigl(\nabla^2_{ij}g\Bigr)^{-1}\nabla_j v$ in Eq.(B2) with its average value:  
\be
\Bigl\langle\nabla_iv\Bigl(\nabla^2_{ij}g\Bigr)^{-1}\nabla_j v \Big\vert \eta_i=0\Bigr\rangle =
\Sigma^2_{\eta_v|\eta}\Bigl(\nabla^2_{ii}g\Bigr)^{-1}
\ee
(recall that $\langle\nabla_i v|\eta_i= 0\rangle = 0$).
So, if $\gamma_{\eta_v\eta}\approx 1$ then this term makes little difference.
In any case, the analysis in the main text ignores this term on the grounds that, for the highest peaks, it matters little.

The main text makes the point that, in addition to maxima of $g_R$, we are also interested in minima of sufficiently high $g_R$.  The statistics of these `high minima' can be written similarly to Eq.~(\ref{fullstats}), except that we must now reverse the signs of the quantities in the two step functions and replace $C_{\rm max}(g_R)\to C_{\rm min}(g_R)$, which requires $4/3<g_R<2.18$.  Whereas symmetry between positive and negative fluctuations in a gaussian field would have the statistics of `high, positive peaks' being similar to those of `low, negative minima' (for which the sign of $g_R$ is also reversed), this will not be true for the `high minima' of interest here, because the signs of the second derivatives of $g_R$ have been reversed, but that of $g_R$ has not.  Generically, we expect the high minima to be less abundant, because $g_R$ is required to be larger, but there will be an additional effect -- for most power spectra a suppression -- due to the difference in sign.  In addition, statistical homogeneity means that large values of $g_R$ become ever rarer as $R$ increases:  i.e., for any $\xo$, $\g\to 0$ for large $R$.  Therefore, if $\xo$ is a peak position which satisfies $C_{\rm max}$ on some $R$, then, although the same position is very unlikely to satisfy either $C_{\rm max}$ or $C_{\rm min}$ on a larger $R$, if it does, we must be careful to only count the position once.

While double-counting is unlikely for high maxima, it is more of a concern for `high minima'.  This is because $g_R$ must eventually turn over and decrease as $R$ increases.  When it turns over, it will have $v_R=0$ and $w_R>0$, but because $g_R$ will be larger than the upper bound for $C_{\rm max}$ it will not form a PBH, so there is no double-counting problem.  However, as $g_R$ is guaranteed to satisfy $C_{\rm max}$ on some even larger $R_p>R$, it may also be both a spatial peak and has $w_R>0$ on this larger $R_p$.  If so, then this position $\xo$ potentially contributes to the PBH counts both on scale $R$ and on scale $R_p$.  To be a maximum on scale $R_p$, it must have been a minimum on some scale between $R$ and $R_p$ before zig-zagging back up at $R_p$.  Clearly, zig-zags are problematic, and they will be more problematic for high minima than for maxima.  We refer readers who are worried about zig-zag related double-counting to \cite{zigzag} for a complete discussion of how to include this, exactly, in the analysis.  However, for most power spectra of cosmological interest, zig-zags are increasingly unlikely at large $R$ (see also \cite{riottocount}).  Since we are interested in the large $R$ regime, we have not included this additional complexity in our analysis, but note that, in principle, we could have done so.

\section{Conditional peaks}\label{condpk}
The same statistics described in the main text might be more intuitively re-written by using the constrained variances $\sigma_{g|v}^2\equiv \sigma_0^2(1 - \gamma_{vg}^2)$ and $\sigma_{w|g}^2\equiv \sigma_w^2(1 - \gamma_{wg}^2)$ denoting the variance of $g$ and $w$ at fixed $v$. Then, it is useful to define
\be
 \nu_{g|v}\equiv \frac{g}{\sigma_0\sqrt{1 - \gamma_{vg}^2}} = \frac{g}{\sigma_{g|v}}.
\ee
This makes
\be
 dg\, dw\, p(g,w,v=0) = \frac{1}{\sqrt{2\pi\sigma_v^2}}\,
      d\nu_{g|v}\, \frac{e^{-\nu_{g|v}^2/2}}{\sqrt{2\pi}}\ 
 dw\,\frac{e^{-(w-\langle w|v=0,g\rangle)^2/2\Sigma_{w|vg}^2}}{\sqrt{2\pi\,\Sigma^2_{w|vg}}}
\ee
with
\be
 \frac{\langle w|v=0,g\rangle}{\sigma_{w|v}} = \gamma_{wg|v}\,\nu_{g|v} 
 \qquad{\rm and}\qquad
 \frac{\Sigma^2_{w|gv}}{\sigma_{w|v}^2} = 1 - \gamma_{wg|v}^2
 \qquad {\rm where}\qquad
 \gamma_{wg|v} \equiv \frac{\gamma_{wg} - \gamma_{wv}\gamma_{vg}}{\sqrt{(1 - \gamma_{wv}^2)(1 - \gamma_{gv}^2)}}\, .
 \ee
Finally, for what is to follow, it is useful to know that 
\be
\frac{\langle w|g,v=0\rangle}{\sigma_w}
 = \frac{d\sigma_{0|v}/d\ln R}{d\sigma_0/d\ln R}\,\nu_{g|v}
                         \,\frac{\sigma_{v}^2}{\sigma_w\sigma_0}.
\ee
Suppose we search for spatial peaks on scale $R$ that have $v_R=0$; i.e., we do not require that $w_R>0$.  The number density of high peaks ($\nu_{g|v}\gg 1$) for these constrained statistics can be derived following the steps laid out in \cite{bbks} and which lead to their Eq.~(4.14):
\be
 n_{\rm pk}(\nu_{g|v}|v=0)\,d\nu_{g|v} \to \left(\frac{\gamma_{\chi|v}\,\nu_{g|v}}{\sqrt{6\pi}\,R\sigma_{1|v}/\sigma_{2|v}}\right)^3 \,\frac{e^{-\nu_{g|v}^2/2}}{\sqrt{2\pi}}\,d\nu_{g|v}
 = \left(\frac{\sigma_{1|v}}{\sqrt{3}\sigma_{0|v}}\right)^3\,\nu_{g|v}^3
 \,\frac{e^{-\nu_{g|v}^2/2}}{(2\pi)^2\,R^3}\,d\nu_{g|v}
\ee
Including the constraint that $w_R>0$ makes this
\be
\frac{dn_{\rm pk}}{d\ln R} \to
  \frac{\langle w|g,v=0\rangle}{\sqrt{2\pi}\sigma_v}\,
  \left(\frac{\gamma_{\chi g|v}\,\nu_{g|v}}{\sqrt{6\pi}\,R\sigma_{1|v}/\sigma_{2|v}}\right)^3 \,\frac{e^{-\nu_{g|v}^2/2}}{\sqrt{2\pi}\sigma_{g|v}}
  = \frac{\gamma_{gv}^{-1}}{\sqrt{2\pi}\sigma_0}\,
  \left(\frac{\sigma_{1|v}}{\sqrt{3}\sigma_{0|v}}\right)^3 \,\nu_{g|v}^3
  \,\frac{e^{-\nu_{g|v}^2/2}}{(2\pi)^2\,R^3}\,\frac{d\nu_{g|v}}{d\ln R}
\ee
This is essentially what leads to the high peaks limit in the main text.

\section{Useful expressions for the case of a broad ${\cal P}(k)$}\label{C}
We are interested in ${\cal P}(k) = A_s\,(kr_p)^p\,\exp(-kr_p)$, for $0\le p\le 4$.  First define 
\be
 I_n(\rho)
  \equiv \int_0^\infty \frac{dx}{x}\,\frac{x^n}{2\pi^2}\,W^2(x)\,\exp(-x/\rho),
\ee
since all the correlators of interest (Eq.~\ref{gammas}) can be obtained from 
\be
\sigma_j^2(R) = \frac{16}{81}\,A_s\,2\pi^2\,\frac{I_{p+4+2j}(R/r_p)}{(R/r_p)^{p}}
 \qquad{\rm and}\quad 
 \frac{d\sigma_j^2}{d\ln R} = \frac{16}{81}\,A_s\,2\pi^2\,\left[\frac{I_{p+5+2j}(R/r_p)}{(R/r_p)^{p+1}} - p\frac{I_{p+4+2j}(R/r_p)}{(R/r_p)^p}\right].
 \ee
For integer $n$, the first few $I_n$ are:  
\begin{align}
 I_4 &=
  \frac{9}{4\pi^2}\, \frac{(2 \rho^2 + 1) \ln(4 \rho^2 + 1) - 4\rho^2}
       {4\rho^2}\\
 I_5 &=
  \frac{9}{4\pi^2}\, \rho \left(\frac{4\rho^2 + 2}{4\rho^2 + 1} - \frac{\ln(4 \rho^2 + 1)}{2\rho^2}\right)\\
 I_6 &= \frac{9}{2\pi^2}\,\rho^2 \left(\frac{8\rho^4 - 6\rho^2 - 1}{[4\rho^2 + 1]^2} + \frac{\ln[4\rho^2+1]}{4\rho^2}\right)\\
 I_7 &= \frac{144}{2\pi^2}\, \rho^7\frac{4\rho^2 + 5}{[4 \rho^2 + 1]^3} \\
 I_8 &= \frac{144}{2\pi^2}\,\rho^8
       \frac{48 \rho^4 + 56 \rho^2 + 35}{[4\rho^2 + 1]^4}\\
 I_9 &= \frac{1152}{2\pi^2}\,\rho^9 \frac{96\rho^6 + 128\rho^4 + 70\rho^2 + 35}{[4 \rho^2 + 1]^5}\\
 I_{10} &= \frac{1728}{\pi^2}\,\rho^{10}
       \frac{640 \rho^8 + 992 \rho^6 + 648 \rho^4 + 210 \rho^2 + 105}{[4\rho^2 + 1]^6}.
\end{align}
For $R/r_p\gg 1$,
\be
 \sigma_0^2 \to \frac{16}{81}\,A_s\,\frac{9}{2},\qquad 
 \sigma_1^2 \to \frac{16}{81}\,A_s\,9\rho^2,\qquad
 \sigma_2^2 \to \frac{16}{81}\,A_s\,108\rho^4 \qquad {\rm if}\quad p=1\\
 \sigma_0^2 \to \frac{16}{81}\,A_s\,27,\qquad 
 \sigma_1^2 \to \frac{16}{81}\,A_s\,540\rho^2,\qquad
 \sigma_2^2 \to \frac{16}{81}\,A_s\,22680\rho^4 \qquad {\rm if}\quad p=4
\ee
The fact that $\sigma^2_j/R^{2j}\to $ constant when $R\gg r_p$ is peculiar to the tophat.  To appreciate its origin, suppose we replace the exponential cutoff on ${\cal P}$ with a sharp cutoff at $k=k_{\rm max}$.  Then, setting $\rho\equiv k_{\rm max}R$ yields
\be
 \sigma_j^2(R) = \int_0^{\rho} \frac{dx}{x}\,x^{4+2j} W^2(x)\,(x/\rho)^p.
\ee
It is easy to check that, at large $\rho$, the quantity $\sigma_j^2/\rho^{2j}$ oscillates weakly around a constant, non-zero value.  Softening the cutoff smears out the oscillations, but does not change the fact that the value is constant at $\rho\gg 1$.  Finally, models with $P(k)=0$ at $kr_p<1$, but $\propto (kr_p)^p$ with $p<0$ for $kr_p>1$, such as those studied by \cite{vicente}, have
\be
 \sigma_j^2(R) = \int_\rho^\infty \frac{dx}{x}\,x^{4+2j} W^2(x)\,(x/\rho)^{p}.
\ee
For these too $\sigma_j^2/\rho^{2j}$ oscillates weakly around a constant, non-zero value at $\rho\gg 1$.

To show that this is a feature of the top-hat smoothing filter, suppose instead that $W(kR) = \exp(-k^2R^2/2)$.  To see how this affects the $\sigma_j$, first consider the case in which we cut-off the power spectrum with a Gaussian:  ${\cal P}(k)\propto (kr_p)^p \exp(-k^2r_p^2)$.  Then, 
\be
 \sigma_j^2(R) 
               = \int \frac{dx}{x}\,x^{4+2j}\, (x/\rho)^p\, e^{- x^2 (1+\rho^2)/\rho^2}
               = \left(\frac{\rho^2}{1+\rho^2}\right)^{(4+2j+p)/2)}\,
                 \frac{\Gamma[(4+2j+p)/2]}{2\rho^p}.
\ee
For $\rho\gg 1$, $\sigma_j^2\propto R^{-p}$ decreases as $R$ increases, and ratios of the $\sigma_j$ are independent of $R$.  The decrease of $\sigma_j$ as $R$ increases persists if the cutoff is exponential ${\cal P}(k)\propto (kr_p)^p \exp(-kr_p)$ rather than Gaussian (this is also analytic, but the exact expression is not so intuitive).  The main point is that this large $R\gg r_p$ behaviour is rather different from that for a tophat.  In particular, in this case, because $\sigma_2/\sigma_1$ tends to a constant, $\sigma_2/R\sigma_1\propto R^{-1}$.  As a result, the peaks theory scaling with $R$ will be very different than it is for a top-hat.  It is common to assert that the details of the smoothing filter do not matter.  The discussion above shows this is simply not true: The top-hat is very different from other filters.  However, for us, the functional form of $W$ is not a choice but dictated by the definition of the compaction function. Thus, the constancy of $\sigma_j/\rho^j$ at large $\rho$, is a physical result.

This matters in the current context because $\sigma_0^2$ independent of $R$, as it is for a top-hat, means that large values of $g_R$ are {\em not} less likely as $R$ increases.  Therefore, one may worry that, for a top-hat smoothing filter, accounting more carefully for zig-zags, following the logic and approximations laid out in \cite{zigzag}, may be necessary.  However, the expressions in the main text (Eqs.~\ref{beta}, \ref{betam}) depend both on $g_R$ and on $w_R$.  At large $R$, $\sigma_w\to\sigma_2\propto \rho^2$, can be large even if the amplitude of $P(k)$ is small.  As a result, the `high peak limit' discussed at the end of Section~V cannot be used.  
This is why, when presenting the numerical results in the main text, we do not assume the high peak limit:  when the full expression (the sum of Eqs.~\ref{beta} and~\ref{betam}) is evaluated numerically, $\beta_\bullet$ stabilises to a constant value, suggesting that double-counting arising from zig-zags is not a severe problem.  

The stabilisation at large $R$ can be understood as follows.  At large $\rho$, all the correlators ($\gamma_{gw}$ etc.) become independent of $R$.  In addition, $\sigma_w\to\sigma_\chi$ and $\gamma_{w\chi}\to 1$.  So the peaks theory term $f(y)\to f(\nu_{w|v})$.  Now, $f(y)$ scales as $y^8$ and $y^3 - 3y$ as $y\to 0$ and $y\to\infty$ respectively.  It is then convenient to think of the integrals over $w$ and $g$ as being over dimensionless $\nu_{g|v}\to\nu_g$ and $\nu_{w|g}\to\nu_w$.  If we write their joint distribution as $p(\nu_{w|v})\,p(\nu_{g|v}|\nu_{w|v})$, and we use the fact that $f(\nu_{w|v})\propto \nu_{w|v}^8$ at small $\nu_{w|v}$, then the $\nu_{w|v}$ integrand is peaked around a characteristic value:  i.e., $w$ is typically proportional to $\sigma_w$ which is large if $A_s\rho^4$ is sufficiently large.  
  The limits on the integral over $\nu_g$ become $(4/3)/\sigma_0$ etc.; i.e., they depend on $A_s$ but not on $R$ (because $\sigma_0\to$ constant).  In particular, they become big as $A_s$ decreases, and this leads to an overall suppression of the PBH abundances at small $A_s$.  Additional dependence on $A_s$ and $R$ comes from the fact that $[{\cal C}(g) - {\cal C}_c(w)]$ must be positive.  As $R$ increases, $\sigma_w$ increases, and this pushes ${\cal C}_c\to 2/3$.  This forces ${\cal C}(g)$ ever closer to its limiting value of $4/3$, thus decreasing the range of $\nu_g$ which can contribute to the integral.  Hence, at large $R$, the abundances are also strongly suppressed.  Finally, there is an overall polynomial in $R$ which comes from $(dR/R) (R_{\rm eq}\sigma_2/R\sigma_1)^3(H_{\rm eq}/H_R) (\sigma_w/\sigma_v)$.  Since $H_{\rm eq}/H_R\propto (R/R_{\rm eq})^2$, and at $\rho\gg 1$, $\sigma_2/R\sigma_1\to$ constant and $\sigma_w/\sigma_v\to \sigma_2/\sigma_1\propto R$, this pre-factor scales as $dR\, R^2$.  This is not enough to win against the suppression coming from the truncated Gaussian distribution of $g$.

\begin{acknowledgments}
	CG would like to thank Vicente Atal, Jaume Garriga, Kazunori Kohri, Antonio Riotto and Chulmoon Yoo for many illuminating discussions. CG would also thank Christian Byrnes and Ilia Musco for comments on the first version of this paper. CG also thanks the Kavli IPMU for its hospitality during the focus week on primordial black holes where this work was presented for the first time. CG is supported by the Ramon y Cajal program and partially supported by the Unidad de Excelencia Maria de Maeztu Grant No. MDM-2014-0369 and the national FPA2016-76005-C2-2-P grants.
\end{acknowledgments}


\end{document}